\documentstyle[epsfig]{mn2e}

\title[High speed photometry of faint Cataclysmic Variables - VII.]
{High speed photometry of faint cataclysmic variables - VII. 
Targets selected from the Sloan Digital Sky Survey and the Catalina
Real-time Transient Survey}
\author[Patrick A.~Woudt et al.]
       {Patrick A.~Woudt$^{1}$\thanks{email: Patrick.Woudt@uct.ac.za},
Brian Warner$^{1,2}$, Deanne de Bud\'e$^1$, Sally Macfarlane$^1$ \newauthor Matthew 
P.E.~Schurch$^{1}$ \& Ewald Zietsman$^{1}$ \\
        $^1$ Astrophysics, Cosmology and Gravity Centre,
        Department of Astronomy, University of Cape Town, Private Bag X3,\\
        Rondebosch 7701, South Africa\\
        $^2$ School of Physics and Astronomy, Southampton University, Highfield, 
        Southampton SO17 1BJ, UK}
\date{Accepted 2011 December 30. Received 2011 December 30; in original
form 2011 October 18}

\begin{document}

\maketitle

\begin{abstract}
We present high speed photometric observations of 20 faint cataclysmic variables, 
selected from the Sloan Digital Sky Survey and Catalina catalogues. Measurements 
are given of 15 new directly measured orbital periods, including four eclipsing dwarf
novae (SDSS0904+03, CSS0826-00, CSS1404-10 and CSS1626-12), two new polars (CSS0810+00 and
CSS1503-22) and two dwarf novae with superhumps in quiescence (CSS0322+02 and CSS0826-00).
Whilst most of the dwarf novae presented here have periods below 2 h, SDSS0805+07 and SSS0617-36 
have relatively long orbital periods of 5.489 and 3.440 h, respectively. The double humped 
orbital modulations observed in SSS0221-26, CSS0345-01, CSS1300+11 and CSS1443-17 are typical of
low mass transfer rate dwarf novae. The white dwarf
primary of SDSS0919+08 is confirmed to have non-radial oscillations and quasi-periodic oscillations
were observed in the short-period dwarf nova CSS1028-08 during outburst.
We further report the detection of a new nova-like variable (SDSS1519+06). 
The frequency distribution of orbital periods of CVs in the Catalina survey 
has a high peak near $\sim 80$ min orbital period, independently confirming that found 
by G\"ansicke et al (2009) from SDSS sources. We also observe a marked correlation between
the median in the orbital period distribution and the outburst class, in the sense that dwarf novae
with a single observed outburst (over the 5-year baseline of the CRTS coverage) 
occur predominantly at shortest orbital period.

\end{abstract}


\section{Introduction}

In six previous papers (references given in Woudt \& Warner 2010) we presented results 
from high speed photometry of faint cataclysmic variables (CVs: see Warner 1995a for a 
review) that had previously only been poorly studied. We have now observed a further 20 stars, 
drawn from the Sloan Digital Sky Survey (SDSS: e.g., Abazajian et al. 2009) and the Catalina 
Real-Time Transient Survey (CRTS: Drake et al. 2009). 

The CRTS has detected $\sim 650$ CVs to date (2011 October) using a network of three telescopes: the
0.7-m Catalina Sky Survey (CSS) Schmidt telescope, the Mt. Lemmon Survey (MLS) 1.5-m Cassegrain 
and the Siding Springs Survey (SSS) 0.5-m Schmidt telescope.
The CRTS keeps variable objects under surveillance for 21 nights of each lunation, 
giving good coverage of long-term light curves, and publishes results as they are 
obtained\footnote{\tt http://nessi.cacr.caltech.edu/catalina/AllCV.html \\
http://nessi.cacr.caltech.edu/MLS/AllCV.html\\
http://nessi.cacr.caltech.edu/SSS/AllCV.html}.
The CVs detected by the CRTS are largely dwarf novae found from their outbursts - their detection
is triggered by outburst amplitudes in excess of 2 mag - and some magnetic CVs (polars) varying between
low and high states. 
Given the depth of the CRTS survey ($V \sim 20.5$ mag), CRTS is able to sample a population
of intrinsically faint dwarf novae.

In our high-speed photometric survey we generally sample 
CRTS CVs accessible to the Southern African Large Telescope (SALT) in case spectroscopic follow-up 
is required. This implies that we primarily select CRTS CVs south of declination +10$^{\circ}$.
Within this declination range there are 357 CVs in the CRTS (at 2011 October), of which 241 have been 
discovered by the CSS telescope, 7 by the MLS telescope (unique identifications, not duplicated in CSS) and 
109 have been discovered solely by the SSS telescope.

     All of our observations were made with the University of Cape Town (UCT) CCD photometer, 
as described by O'Donoghue (1995), in frame transfer mode and with white light, on the 1.9-m 
(74-in) and 1.0-m (40-in) reflectors at the Sutherland site of the South African Astronomical 
Observatory (SAAO). 

     In our previous papers, an approximate magnitude scale was derived with the use of hot white dwarf 
standards (Landolt 1992), but because of the non-standard spectral distributions of CVs and the use of white 
light, our magnitudes approximate a $V$ scale only to $\sim$0.1 mag. Given that our targets in this paper are
largely drawn from the SDSS and CRTS surveys, we explored the reliability of using SDSS photometry
to calibrate our white light observations. On clear nights when hot white dwarf standards were
observed, SDSS stars in the field of view of target CVs revealed a stable zero point offset between
$V$ and SDSS \emph{r} of 0.12 $\pm$ 0.05 over a broad range of SDSS $({g-r})$ colour from $0.2 - 1.0$. 
This is consistent with the SDSS {\em r} to $V$ photometric transformation described by 
Jester et al.~(2005). The UCT CCD magnitudes quoted in this paper correspond to \emph{r} photometric
system, calibrated by SDSS photometry of stars in the field of view of our targets with $({g-r})$ 
colours in the range 0.2 -- 1.0.

Our work is an ongoing survey -- many of the stars reported here will require more intensive 
and extended study. In Section 2 we give the results of our observations. Section 3 gives brief conclusions.

\section{Observations}

\begin{table*}
 \centering
  \caption{Observing log.}
  \begin{tabular}{@{}llrrrrrcc@{}}
 Object            & Type  & Run No.  & Date of obs.          & HJD of first obs. & Length    & $t_{in}$ & Tel. &  \emph{r} \\
                   &       &          & (start of night)      &  (+2450000.0)     & (h)       &     (s)   &      & (mag) \\[10pt]
{\bf SDSS0805+07}  & CV    & S7767 & 2007 Dec 29 & 4464.42448 & 4.10 &  60 & 40-in & 17.9 \\
                   &       & S7771 & 2007 Dec 31 & 4466.52235 & 1.77 &  60 & 40-in & 17.9 \\
                   &       & S7773 & 2008 Jan 01 & 4467.38970 & 4.92 &  60 & 40-in & 17.9 \\[5pt]
{\bf SDSS0904+03}  & CV    & S7190 & 2003 Dec 23 & 2997.53850 & 1.53 &  80 & 74-in & 19.3$^*$ \\
eclipsing          &       & S7200 & 2003 Dec 27 & 3001.49672 & 2.65 &  90 & 74-in & 19.4$^*$ \\
                   &       & S7216 & 2003 Dec 31 & 3005.41754 & 2.91 &  45 & 74-in & 19.2$^*$ \\
                   &       & S7541 & 2004 Dec 14 & 3354.50212 & 2.29 &  80 & 40-in & 19.0$^*$ \\
                   &       & S7545 & 2004 Dec 16 & 3356.50347 & 2.40 &  80 & 40-in & 19.0$^*$ \\
                   &       & S7600 & 2005 Mar 08 & 3438.26907 & 3.52 & 100 & 40-in & 19.2$^*$ \\
                   &       & S7602 & 2005 Mar 09 & 3439.26052 & 2.53 & 100 & 40-in & 19.2$^*$ \\
                   &       & S7603 & 2005 Mar 10 & 3440.26381 & 2.36 & 100 & 40-in & 19.2$^*$ \\
                   &       & S7604 & 2005 Mar 11 & 3441.26385 & 2.83 & 100 & 40-in & 19.2$^*$ \\
                   &       & S7612 & 2005 Mar 29 & 3459.23798 & 2.12 &  60 & 40-in & 19.2$^*$ \\[5pt]
{\bf SDSS0919+08}  & CV    & S7212 & 2003 Dec 30 & 3004.47499 & 2.15 &  30 & 74-in & 18.2:\\
                   &       & S7219 & 2004 Jan 01 & 3006.43116 & 2.29 &  25 & 74-in & 18.3 \\
                   &       & S7235 & 2004 Feb 09 & 3045.34260 & 1.83 & 100 & 40-in & 18.2 \\
                   &       & S7236 & 2004 Feb 13 & 3049.36907 & 0.48 & 100 & 40-in & 18.1 \\
                   &       & S7238 & 2004 Feb 14 & 3050.32530 & 2.80 &  90 & 40-in & 18.2 \\
                   &       & S7243 & 2004 Feb 15 & 3051.33939 & 2.24 & 100 & 40-in & 18.2 \\
                   &       & S7561 & 2005 Feb 01 & 3403.44323 & 2.95 &  90 & 40-in & 18.3 \\
                   &       & S7564 & 2005 Feb 03 & 3405.40375 & 2.55 &  90 & 40-in & 18.1:\\
                   &       & S7566 & 2005 Feb 04 & 3406.39126 & 4.05 &  90 & 40-in & 18.3 \\
                   &       & S7569 & 2005 Feb 05 & 3407.39868 & 3.47 &  90 & 40-in & 18.2 \\
                   &       & S7572 & 2005 Feb 06 & 3408.41099 & 2.73 &  90 & 40-in & 18.2 \\
                   &       & S7575 & 2005 Feb 07 & 3409.40965 & 1.60 &  90 & 40-in & 18.2 \\
                   &       & S7685 & 2007 Jan 14 & 4115.42552 & 3.83 &  30 & 40-in & 18.3 \\
                   &       & S7687 & 2007 Jan 15 & 4116.42147 & 4.11 &  30 & 40-in & 18.3 \\
                   &       & S7689 & 2007 Jan 16 & 4117.43745 & 2.91 &  30 & 40-in & 18.2 \\
                   &       & S7692 & 2007 Jan 17 & 4118.40030 & 4.93 &  30 & 40-in & 18.2 \\
                   &       & S7695 & 2007 Jan 18 & 4119.41123 & 4.33 &  30 & 40-in & 18.2 \\
                   &       & S7698 & 2007 Jan 19 & 4120.42478 & 4.14 &  30 & 40-in & 18.2 \\
                   &       & S7702 & 2007 Jan 21 & 4122.42529 & 4.27 &  30 & 40-in & 18.2 \\
                   &       & S7705 & 2007 Jan 22 & 4123.44064 & 3.66 &  30 & 40-in & 18.2 \\
                   &       & S7707 & 2007 Jan 23 & 4124.40915 & 3.83 &  30 & 40-in & 18.2 \\[5pt]
{\bf SDSS1519+06}  & CV    & S7759 & 2007 Mar 24 & 4184.54459 & 2.77 &  40 & 40-in & 17.1$^b$ \\
                   &       & S7762 & 2007 Mar 26 & 4186.48546 & 4.14 &  40 & 40-in & 17.2$^b$ \\
                   &       & S7765 & 2007 Mar 27 & 4187.47591 & 4.11 &  40 & 40-in & 17.0$^b$ \\[5pt]
{\bf SSS0221-26}   & DN    & S8001 & 2010 Oct 28 & 5498.41592 & 2.34 &  45 & 74-in & 19.3: \\
                   &       & S8004 & 2010 Oct 30 & 5500.33806 & 4.85 &  60 & 74-in & 19.3: \\
                   &       & S8006 & 2010 Oct 31 & 5501.42365 & 2.42 &  60 & 74-in & 19.3: \\[5pt]
{\bf CSS0332+02}   & DN    & S7881 & 2009 Dec 16 & 5182.29440 & 3.98 &  90 & 74-in & 20.1 \\
                   &       & S7883 & 2009 Dec 17 & 5183.28893 & 3.30 &  90 & 74-in & 20.1 \\
                   &       & S7885 & 2009 Dec 18 & 5184.29130 & 3.13 &  90 & 74-in & 20.3 \\ 
                   &       & S7889 & 2009 Dec 20 & 5186.29202 & 3.07 &  90 & 74-in & 20.3 \\
                   &       & S7894 & 2009 Dec 22 & 5188.35299 & 1.48 &  90 & 74-in & 20.0 \\
                   &       & S7896 & 2009 Dec 23 & 5189.29053 & 2.85 &  90 & 74-in & 19.9 \\[5pt]
{\bf CSS0334-07}   & DN    & S7680 & 2007 Jan 12 & 4113.28517 & 1.92 &  10 & 40-in & 17.8 \\
                   &       & S7887 & 2009 Dec 19 & 5185.29124 & 3.18 &  30 & 74-in & 18.3 \\
                   &       & S7996 & 2010 Sep 19 & 5458.52966 & 2.88 &  95 & 40-in & 18.4 \\
                   &       & S8009 & 2010 Nov 01 & 5502.42705 & 4.63 &  45 & 74-in & 18.6 \\
                   &       & S8043 & 2010 Dec 12 & 5543.29738 & 3.80 &  60 & 74-in & 18.6 \\[5pt]
{\bf CSS0345-01}   & DN    & S8016 & 2010 Nov 26 & 5527.29721 & 4.32 &90,100& 40-in & 18.6: \\
                   &       & S8019 & 2010 Nov 27 & 5528.39225 & 1.70 &  90 & 40-in & 18.6: \\
                   &       & S8026 & 2010 Nov 30 & 5531.34710 & 5.18 &  90 & 40-in & 18.7: \\[5pt]

\end{tabular}
{\footnotesize 
\newline 
Notes: DN = Dwarf Nova; $t_{in}$ is the integration time; $^*$ mean magnitude out of eclipse; `:' denotes an uncertain value; $^b$ observations taken with the SAAO CCD.\hfill}
\label{survey7tab1}
\end{table*}
\addtocounter{table}{-1}

\begin{table*}
 \centering
  \caption{Continued: Observing log.}
  \begin{tabular}{@{}llrrrrrcc@{}}
 Object            & Type  & Run No.  & Date of obs.          & HJD of first obs. & Length    & $t_{in}$ & Tel. &  {\em r} \\
                   &       &          & (start of night)      &  (+2450000.0)     & (h)       &     (s)   &      & (mag) \\[10pt]
{\bf SSS0617-36}   & CV    & S8013 & 2010 Nov 24 & 5525.39167 & 2.23 &  30 & 40-in & 17.6 \\
                   &       & S8015 & 2010 Nov 25 & 5526.39259 & 3.45 &40,45& 40-in & 17.7 \\
                   &       & S8017 & 2010 Nov 26 & 5527.48668 & 2.52 &  40 & 40-in & 17.7 \\
                   &       & S8020 & 2010 Nov 27 & 5528.48484 & 2.91 &  30 & 40-in & 17.7 \\
                   &       & S8029 & 2010 Dec 03 & 5534.35377 & 5.66 &  20 & 40-in & 17.9 \\[5pt]
{\bf CSS0810+00}   & Polar & S7915 & 2010 Mar 18 & 5274.25067 & 4.21 &  20 & 74-in & 18.1 \\
                   &       & S7917 & 2010 Mar 19 & 5275.24505 & 2.81 &  20 & 74-in & 18.3 \\
                   &       & S7923 & 2010 Mar 21 & 5277.24549 & 3.15 &  20 & 74-in & 18.3 \\
                   &       & S7937 & 2010 Apr 02 & 5289.25213 & 2.03 &  60 & 40-in & 18.2 \\[5pt]
{\bf CSS0826-00}   & DN    & S7892 & 2009 Dec 21 & 5187.44515 & 3.60 &  90 & 74-in & 20.1$^*$ \\
eclipsing          &       & S7943 & 2010 Apr 06 & 5293.23279 & 3.49 &  90 & 40-in & 19.7$^*$ \\
                   &       & S7946 & 2010 Apr 07 & 5294.22280 & 3.42 &  90 & 40-in & 19.9$^*$ \\
                   &       & S7957 & 2010 Apr 11 & 5298.21867 & 2.13 &  90 & 40-in & 19.9$^*$ \\[5pt]
{\bf CSS1028-08}   & DN    & S7918 & 2010 Mar 19 & 5275.40553 & 2.41 &   6 & 74-in & 16.1 \\
                   &       & S7921 & 2010 Mar 20 & 5276.38365 & 3.87 &10,15& 74-in & 17.5 \\
                   &       & S7924 & 2010 Mar 21 & 5277.39139 & 4.25 &  20 & 74-in & 19.0 \\[5pt]
{\bf CSS1300+11}   & DN    & S7971 & 2010 May 21 & 5338.31041 & 3.15 &  55 & 74-in & 19.9 \\
                   &       & S7987 & 2010 Jun 05 & 5353.21030 & 4.14 &  65 & 74-in & 19.7 \\
                   &       & S7989 & 2010 Jun 06 & 5354.22694 & 2.41 &  65 & 74-in & 19.8 \\
                   &       & S7991 & 2010 Jun 07 & 5355.21884 & 2.22 &  65 & 74-in & 19.6 \\[5pt]
{\bf CSS1321+01}   & DN    & S7935 & 2010 Apr 01 & 5288.43811 & 3.39 & 100 & 40-in & 19.4 \\
{\bf = HV Vir}     &       & S7939 & 2010 Apr 02 & 5289.48174 & 3.45 & 100 & 40-in & 19.5 \\
                   &       & S7944 & 2010 Apr 06 & 5293.38817 & 4.17 & 100 & 40-in & 19.5 \\
                   &       & S7947 & 2010 Apr 07 & 5294.37465 & 4.47 & 100 & 40-in & 19.4 \\
                   &       & S7950 & 2010 Apr 08 & 5295.37400 & 1.61 & 100 & 40-in & 19.5 \\[5pt]
{\bf CSS1404-10}   & DN    & S7833 & 2009 Feb 27 & 4890.58682 & 1.00 &  60 & 74-in & 19.6$^*$ \\
eclipsing          &       & S7848 & 2009 Mar 20 & 4911.45607 & 3.75 &  15 & 74-in & 16.6$^*$ \\
                   &       & S7851 & 2009 Mar 21 & 4912.46066 & 4.76 &  15 & 74-in & 17.3$^*$ \\
                   &       & S7853 & 2009 Mar 22 & 4913.43818 & 2.15 &  30 & 74-in & 18.4$^*$ \\
                   &       & S7858 & 2009 Mar 23 & 4914.48481 & 4.16 &  30 & 74-in & 18.8$^*$ \\
                   &       & S7862 & 2009 Jun 12 & 4995.21939 & 2.63 &  45 & 74-in & 19.4$^*$ \\[5pt]
{\bf CSS1443-17}   & DN    & S7865 & 2009 Jun 17 & 5000.19997 & 4.28 &  60 & 74-in & 19.1 \\
                   &       & S7869 & 2009 Jun 18 & 5001.20840 & 3.50 &  60 & 74-in & 19.1 \\[5pt]
{\bf CSS1503-22}   & Polar & S7925 & 2010 Mar 21 & 5277.58370 & 1.78 &  90 & 74-in & 17.2 \\
                   &       & S7929 & 2010 Mar 22 & 5278.60089 & 1.57 &  55 & 74-in & 17.2 \\
                   &       & S7931 & 2010 Mar 23 & 5279.52952 & 3.01 &  20 & 74-in & 17.5 \\
                   &       & S7962 & 2010 Apr 12 & 5299.48901 & 4.05 &  30 & 40-in & 17.2 \\
                   &       & S7965 & 2010 Apr 13 & 5300.45963 & 4.50 &  30 & 40-in & 17.3 \\[5pt]
{\bf CSS1528+03}   & DN    & S7940 & 2010 Apr 02 & 5289.63048 & 0.60 &  20 & 40-in & 17.0 \\
                   &       & S7956 & 2010 Apr 10 & 5297.45581 & 4.40 &  90 & 40-in & 19.5 \\
                   &       & S7959 & 2010 Apr 11 & 5298.41794 & 5.73 &  90 & 40-in & 19.3 \\
                   &       & S7961 & 2010 Apr 12 & 5299.41131 & 1.70 &  90 & 40-in & 19.9 \\[5pt]
{\bf CSS1626-12}   & DN    & S7985 & 2010 Jun 04 & 5352.40913 & 4.43 &  90 & 74-in & 20.4$^*$ \\
eclipsing          &       & S7988 & 2010 Jun 05 & 5353.40403 & 3.95 &  90 & 74-in & 20.3$^*$ \\
                   &       & S7990 & 2010 Jun 06 & 5354.41289 & 3.05 &  90 & 74-in & 20.4$^*$ \\[5pt]
{\bf CSS2325-08}   & DN    & S7814 & 2008 Oct 15 & 4755.26078 & 1.90 &  60 & 74-in & 19.3:\\
{\bf = EG Aqr}     &       & S7815 & 2008 Oct 16 & 4756.25083 & 4.58 &  30 & 74-in & 19.3:\\
                   &       & S7817 & 2008 Oct 17 & 4757.24637 & 5.50 &  60 & 74-in & 19.3:\\
                   &       & S7819 & 2008 Oct 18 & 4758.24330 & 6.31 &  60 & 74-in & 19.2:\\
                   &       & S7821 & 2008 Oct 19 & 4759.25972 & 4.02 &  60 & 74-in & 19.3:\\
                   &       & S7824 & 2008 Oct 20 & 4760.38167 & 1.22 &  60 & 74-in & 19.4:\\[5pt]
\end{tabular}
{\footnotesize 
\newline 
Notes: DN = Dwarf Nova; $t_{in}$ is the integration time; $^*$ mean magnitude out of eclipse; `:' denotes an uncertain value.\hfill}
\end{table*}

\subsection{SDSS0805+07 (SDSS\,J080534.49+072029.1)}

SDSS0805+07 was announced as a CV in paper VI of the Sloan Digital 
Sky Survey series (Szkody et al.~2007) where it is listed at magnitude 
$r = 17.79$ and $(g-r)$ = 0.73. The SDSS spectrum contained signs of an early type K absorption 
spectrum, suggesting that the orbital period would be quite long. The later 
paper by G\"ansicke et al.~(2009) on orbital periods of SDSS CVs does not 
include any further observations of this star.

\begin{figure}
\centerline{\hbox{\psfig{figure=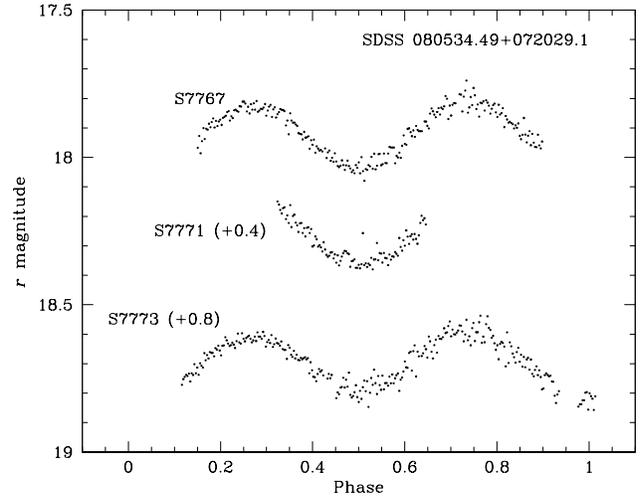,width=8.4cm}}}
  \caption{Individual light curves of SDSS 0805, folded on the 
ephemeris given in Eq.~\ref{ephsdss0805}. The light curve of S7767 is
displayed at the correct brightness. Runs S7771 and S7773 have been
displaced vertically by 0.4 and 0.8 mag, respectively.}
 \label{sdss0805fig1}
\end{figure}

\begin{figure}
\centerline{\hbox{\psfig{figure=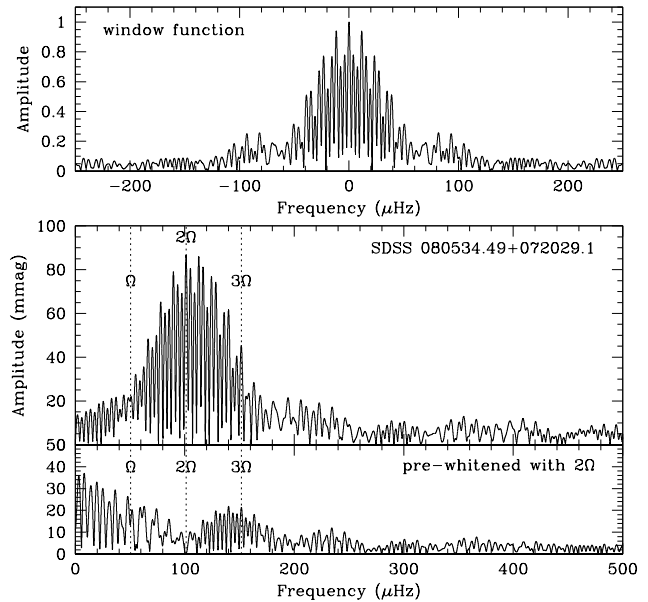,width=8.4cm}}}
  \caption{The top panel displays the window function of the FT of the
combined runs S7767, S7771 and S7773. The middle panel shows the FT of
the combined runs of SDSS 0805 with the orbital frequency ($\Omega$) and its first
two harmonics marked by the dashed vertical lines. The lower panel shows the 
combined FT pre-whitened at the first harmonic ($2\Omega$) of the orbital frequency.}
 \label{sdss0805fig2}
\end{figure}

   Our photometric observations of SDSS0805+07 are listed in 
Tab.~\ref{survey7tab1} and the light curves are shown in 
Fig.~\ref{sdss0805fig1}. The Fourier transform of the data is 
shown in Fig.~\ref{sdss0805fig2} where we have identified the highest
peak as twice the orbital frequency ($2 \Omega$, where $\Omega = 1/P_{orb}$). 
We chose this because of the presence of the significant $3 \Omega$ signal and a weak $\Omega$ 
signal which shows in the light curve as a slightly larger amplitude for
alternate maxima. This choice of orbital period is also driven by the
K absorption spectrum mentioned above; the light curve is evidently 
substantially affected by the ellipsoidal variation of the secondary.
The ephemeris for minimum 
light is

\begin{equation}
    {\rm HJD_{min}} = 245\,4464.39 + 0.2287\,(5)\, {\rm E.}
\label{ephsdss0805}
\end{equation}

\subsection{SDSS0904+03 (SDSS\,J090403.49+035501.2)}

\begin{figure}
\centerline{\hbox{\psfig{figure=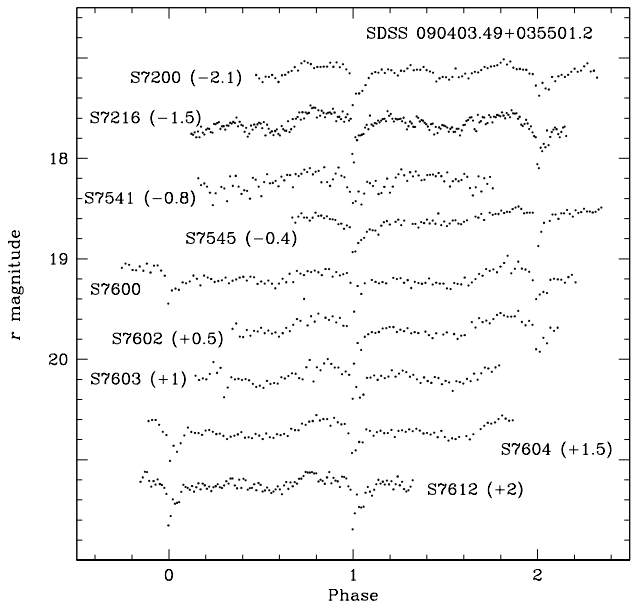,width=8.4cm}}}
  \caption{Individual light curves of SDSS\,0904, 
folded on the ephemeris given in Eq.~\ref{ephsdss0904}. The light curve
of S7600 is displayed at the correct brightness; vertical offsets for each
light curve are given in brackets.}
 \label{sdss0904fig1}
\end{figure}

\begin{figure}
\centerline{\hbox{\psfig{figure=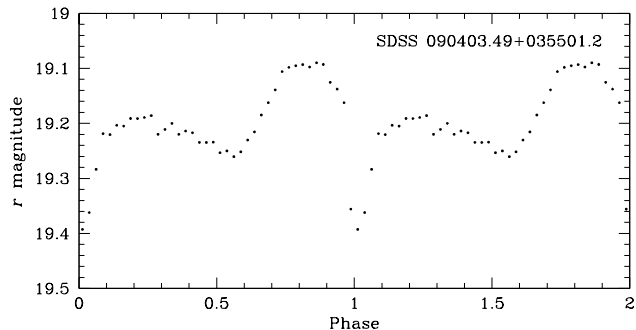,width=8.4cm}}}
  \caption{The average light curve of SDSS\,0904 during March 2005, 
folded on the ephemeris given in Eq.~\ref{ephsdss0904}.}
 \label{sdss0904fig2}
\end{figure}

SDSS0904+03 was listed as a CV in paper III of the 
SDSS series (Szkody et al.~2004) where it was shown to be a 
shallow eclipsing system with $P_{orb} = 85.98$ min and 
prominent white dwarf absorption lines evident in the spectrum. 
Later photometry by Woudt et al.~(2005) showed possible low amplitude 
(24 mmag) variations with a period $\sim 740$ s, classifying it 
as a candidate CV/ZZ star - i.e. a CV containing a nonradially pulsating white 
dwarf (a ZZ Cet star). Our observations are listed in Tab.~\ref{survey7tab1}, 
the individual light curves are shown in Fig.~\ref{sdss0904fig1} and 
an average light curve for the March 2005 runs is shown in 
Fig.~\ref{sdss0904fig2}, which is typical of a low mass transfer
dwarf nova. An FT of the whole data set 
provides $P_{orb} = 86.0015$ min and the eclipse ephemeris  

\begin{equation}
    {\rm HJD_{min}} = 245\,3438.2843 + 0.05972325\,(2)\, {\rm E.}
\label{ephsdss0904}
\end{equation}

There are no significant peaks in the FTs at higher frequencies in
our larger data set, so SDSS0904+03 appears not to be a pulsating
system. We now think that the earlier tentative detection of power
at 740 s was caused by a high harmonic of the eclipse profile.

\subsection{SDSS0919+08 (SDSS\,J091945.11+085710.0)}

SDSS0919+08 was recognised as a CV in paper IV of the 
SDSS series (Szkody et al.~2005), where it is given as magnitude 
$g = 18.20$. The spectrum shows Balmer emission with underlying 
absorption spectrum of a white dwarf. From a small number of 
spectra an orbital period of $84 \pm 8$ min was deduced. 
Mukadum et al.~(2007) detected a periodic signal in the light 
curve which gave $91 \pm 7$ min for $P_{orb}$ and found a 
further low amplitude ($\sim 1$\%) brightness modulation 
near 260 s in 5 out of 6 light curves, which might be a 
doublet with splitting $\sim 1.3$ s. They estimated an effective 
temperature of 13\,000 K for the white dwarf. The nature of the 260 s 
periodicity was not ascertained - it could be a rotationally or 
magnetically split nonradial oscillation, or directly related 
to a spin period.

   Dillon et al.~(2008) found an orbital period of $81.6 \pm 1.2$ min 
from a 3 h high speed photometry run on one night with the Calar Alto 
2.2-m reflector, and Szkody et al.~(2010) found a possible $P_{orb}$ harmonic 
at 40.75 min with the Hubble Space Telescope but no high frequency modulation at all with 
various ground based telescopes in 2007/2008. Their conclusion was that 
SDSS0919+08 had at least temporarily stopped pulsating.

\begin{figure}
\centerline{\hbox{\psfig{figure=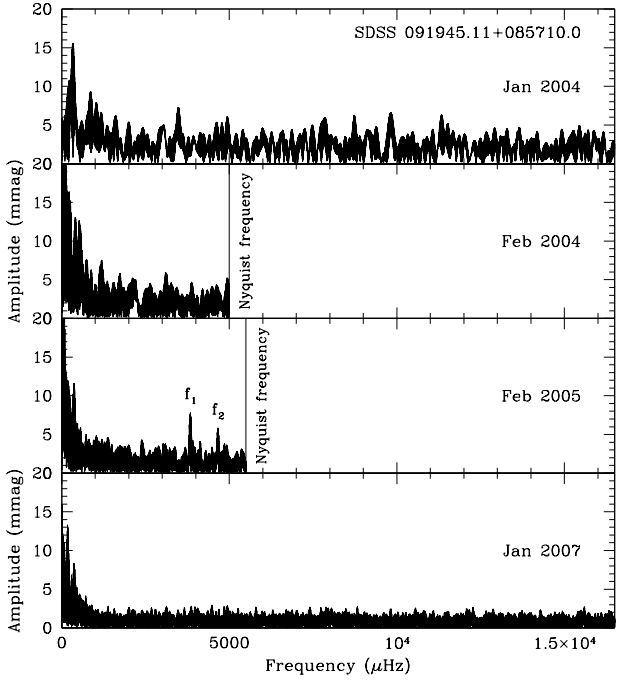,width=8.4cm}}}
  \caption{The combined FTs of SDSS\,0919 during 2004 January, February, 2005 February and 2007 
January, respectively.}
 \label{sdss0919fig1}
\end{figure}

\begin{figure}
\centerline{\hbox{\psfig{figure=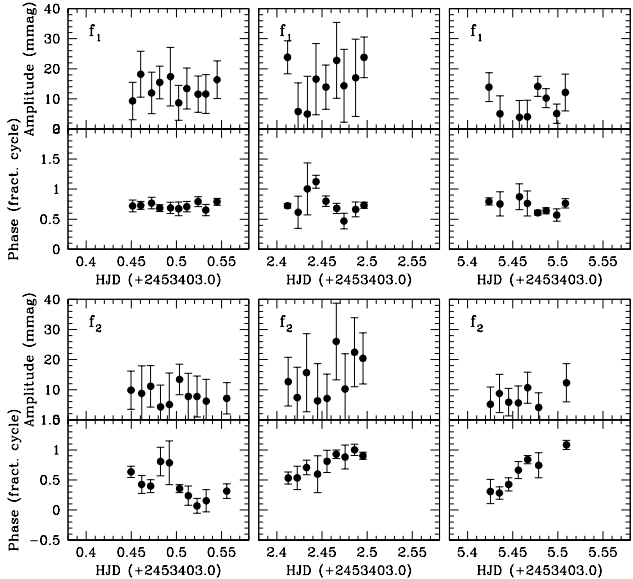,width=8.4cm}}}
  \caption{The phase ($\phi$) and amplitude ($A$) diagrams 
for $f_1$ (upper two rows) and $f_2$ (bottom two rows) of SDSS0919 on three 
different nights in 2005 February. Each dot represents $\sim 5$ cycles of the 
260-s modulation ($f_1$) or $\sim 6$ cycles of the 214-s modulation ($f_2$), with a 
33\% overlap. In the phase diagrams, only those points are plotted where the 
amplitude of the modulation is larger than 3 mmag.}
 \label{sdss0919fig2}
\end{figure}

   Our photometric runs on SDSS0919+08 are listed in Tab.~\ref{survey7tab1}; 
they were made mostly in 2004 February, 2005 February and 2007 January. 
The FTs are shown in Fig.~\ref{sdss0919fig1}. 
We saw no modulation above 5\% in 2004 and none above 3\% in 2007 
(in agreement with Szkody et al.~observations described above) but in 
2005 February there was not only a $\sim 8$\% modulation at 260 s but 
also a 6\% amplitude at 214 s. The first of these is a $\sim 3 \sigma$ 
spike and is at the period often seen before, so there is no doubt about 
its reality. For the shorter period oscillation we show in Fig.~\ref{sdss0919fig2} an
amplitude ($A$) - phase ($\phi$) diagram for runs S7561, S7564 and S7572, respectively,
obtained in 2005 February. These illustrate the high coherence of the 214-s oscillation
($f_2$) with a drift in frequency that is commonly seen in oscillation white dwarfs with
unresolved multiplets.

The presence of two simultaneous frequencies excludes a simple 
white dwarf rotation model for SDSS\,J0919+08, but is indicative of
nonradial oscillations in the white dwarf.

\subsection{SDSS1519+06 (SDSS\,J151915.86+064529.1)}

\begin{figure}
\centerline{\hbox{\psfig{figure=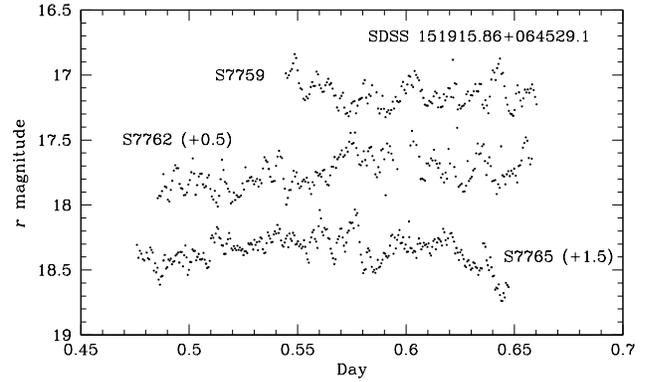,width=8.4cm}}}
  \caption{Individual light curves of SDSS\,1519+06.  The light curve
of run S7759 is displayed at the correct brightness; vertical offsets of each
light curve are given in brackets.}
 \label{sdss1519fig1}
\end{figure}

SDSS1519+06 is another object in the Sloan data base, listed 
at $r = 16.74$ and $(g-r) = -0.17$, which has colours similar to that 
of CVs. We observed it on three nights 
in 2007 March (see Tab.~\ref{survey7tab1}) using the SAAO CCD; the light 
curves are displayed in Fig.~\ref{sdss1519fig1} and show very active 
flickering characteristic of a CV but no periodic modulations.

Since making these observations an SDSS spectrum is available on the SDSS
Skyserver which shows a strong blue continuum, narrow Balmer emission lines and a strong
He\,II emission which are characteristic of a low inclination nova-like variable.

\subsection{SSS0221-26 (SSS100511:022138-261952)}

\begin{figure}
\centerline{\hbox{\psfig{figure=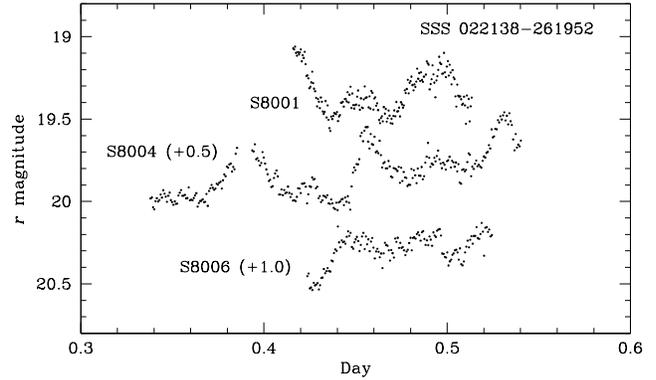,width=8.4cm}}}
  \caption{Individual light curves of SSS0221-26.  The light curve
of run S8001 is displayed at the correct brightness; vertical offsets
of each light curve are given in brackets.}
 \label{sss0221fig1}
\end{figure}

SSS0221-26 is a Catalina source (100511:022138-261952)
discovered at the Siding Spring Observatory in Australia. We observed
it on three nights as listed in Tab.~\ref{survey7tab1} and the light
curves are shown in Fig.~\ref{sss0221fig1}.

\begin{figure}
\centerline{\hbox{\psfig{figure=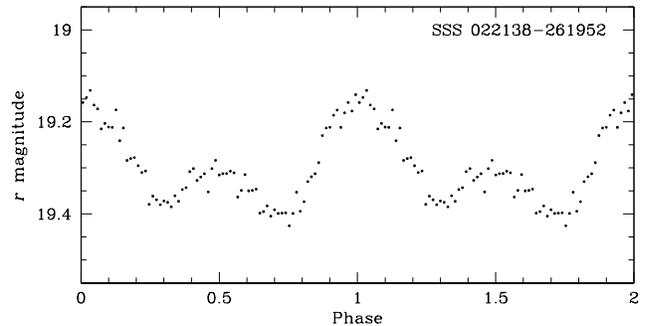,width=8.4cm}}}
  \caption{The average light curve of SSS0221-26, folded on the 
ephemeris given in Eq.~\ref{ephsss0221}.}
 \label{sss0221fig2}
\end{figure}

The FT shows maximum power at the fundamental and first harmonic 
of a 101.5-min periodicity, but the 1-d alias at 109.1 min cannot be excluded. 
The mean light curve at 101.5 min is shown in Fig.~\ref{sss0221fig2} which is again
characteristic of a low mass transfer dwarf nova in quiescence. The ephemeris
for maximum light is:

\begin{equation}
    {\rm HJD_{max}} = 245\,5458.4170 + 0.07049\,(9)\, {\rm E.}
\label{ephsss0221}
\end{equation}

\subsection{CSS0332+02 (CSS091121:033232+020439)}

\begin{figure}
\centerline{\hbox{\psfig{figure=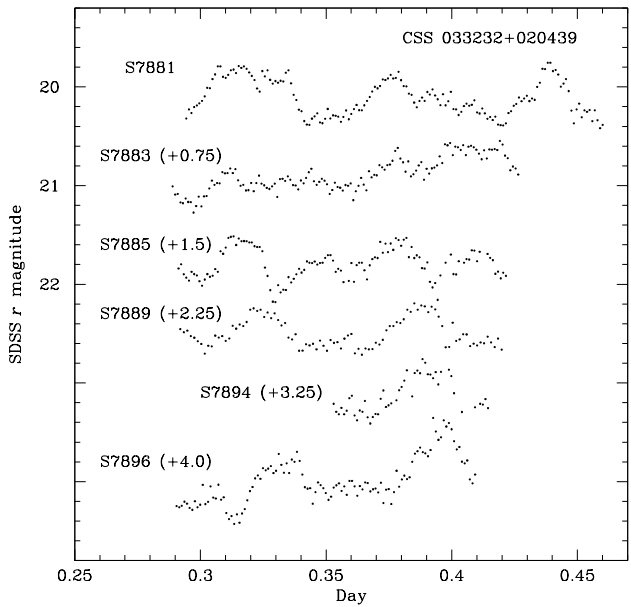,width=8.4cm}}}
  \caption{Individual light curves of CSS0332+02.  The light curve
of run S7881 is displayed at the correct brightness; vertical offsets
of each light curve are given in brackets.}
 \label{css0332fig1}
\end{figure}

\begin{figure}
\centerline{\hbox{\psfig{figure=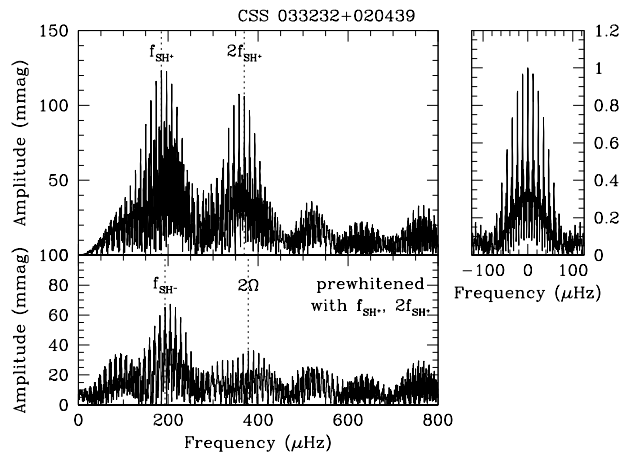,width=8.4cm}}}
  \caption{The Fourier transform of CSS0332+02 for the complete data
set (top panel: left); the mean and trend has been removed for each individual run.
The window function is shown in the top right panel. The lower panel shows the
Fourier transform pre-whitened at the negative superhump frequency and its first
harmonic. Frequencies listed in Tab.~\ref{survey7tab2} have been marked by dashed
vertical lines and labelled. }
 \label{css0332fig2}
\end{figure}

CSS0332+02 is a Catalina source (091121:033232+020439) that has had 
three known outbursts, reaching $V \sim 17$, in 2006 October, 2007 October
and 2009 November. At quiescence it is fainter than $V \sim 20$ in the 
Catalina light curve. Our observations are listed in Tab.~\ref{survey7tab1} 
and were obtained less than a month after the last recorded outburst, 
when CSS0332+02 was at $r \sim 20.1$. The light curves are shown in 
Fig.~\ref{css0332fig1} and exhibit profiles typical of orbital or 
superhump modulations.  They turn out be a combination of both. The FT 
of the full data set is shown in the upper left panel of Fig.~\ref{css0332fig2} - with 
the window pattern next to it. The highest peak in the alias pattern 
in the FT is marked $f_{{sh}^+}$ and its first harmonic is marked  
2$f_{{sh}^+}$; these are the only two components of the two patterns 
that have a 2:1 frequency ratio within the small errors involved 
($\sim 1 \mu$Hz). Pre-whitening with these two frequencies leaves 
the FT shown in the lower panel of Fig.~\ref{css0332fig2} - revealing 
a further alias pattern (which was easily seen in the original total FT), 
where we have marked the maximum component $f_{{sh}^-}$. In addition 
there is a very low amplitude residual alias pattern 
labelled $2 \Omega$. The frequencies are given in Tab.~\ref{survey7tab2}.

   With this choice the implied value of the orbital frequency 
$\Omega = 189.1$ $\mu$Hz is roughly midway between the frequencies 
$f_{{sh}^+}$ and $f_{{sh}^-}$, which we claim are positive 
and negative superhumps. Note that the positive superhump 
excess $\epsilon^+ = ({P_{{sh}^+}}/{P_{orb}}) - 1 =  0.0231$ and 
the negative superhump deficiency $\epsilon^- = ({P_{{sh}^-}}/{P_{orb}}) - 1
= -0.0204$ are close to values for an SU UMa type dwarf nova with 
$P_{orb} \sim 88.14$ min (Patterson et al.~2005). Note also that the 
largest modulation is the positive superhump, and its strong 
first harmonic is typical of the non-sinusoidality of such 
superhumps.   

\begin{table}
 \centering
  \caption{Frequencies observed in CSS0332+02.}
  \begin{tabular}{@{}cccc@{}}
 ID   & Frequency  & Period & Ampl.    \\
      & ($\mu$Hz)  & (h)    & (mag)   \\[10pt]
$f_{sh^+}$     & 184.81 $\pm$ 0.03 & 1.503 &  0.119 $\pm$ 0.006  \\
$2f_{sh^+}$    & 369.27 $\pm$ 0.04 & 0.752 &  0.104 $\pm$ 0.006  \\
$f_{sh^-}$     & 193.08 $\pm$ 0.05 & 1.439 &  0.067 $\pm$ 0.005  \\
$2\Omega$     & 378.23 $\pm$ 0.11 & 0.734 &  0.036 $\pm$ 0.005  \\
\end{tabular}
\label{survey7tab2}
\end{table}

Our interpretation of this data set is that CSS0332+02 is a dwarf 
nova showing both positive and negative superhumps (and the 
first harmonic of $P_{orb}$) in quiescence. We arrived at this 
surprising conclusion partly with the help of the work of Olech, Rutkowski \& 
Schwarzenberg-Czerny (2009) who found that SDSS2100+00 has 
persistent negative superhumps in quiesence, and moreover was the third such 
dwarf nova ``bihumper'' to be found, the others being V503 Cyg 
(Harvey et al.~1995) and BF Ara (Olech, Rutkowski \& Schwarzenberg-Czerny 
2007). But in SDSS2100+00, V503 Cyg and BF Ara the positive superhumps 
are only seen during superoutbursts. The uniqueness of this interpretation
encourages further extensive observations.

\subsection{CSS0334-07 (CSS091027:033450-071048)}

CSS0334-07 is CSS091027:033450-071048 with a quiescent magnitude
$V \sim 18.5$; the Catalina light curve shows outbursts to $V \sim 15$
on a time scale of $\sim$200 d. This star is also listed in the Sloan
Survey as SDSS\,J033449.87-071047.8 (Szkody et al. 2007), identified
as a dwarf nova with $g = 14.64$ (caught by SDSS in outburst) 
with a candidate orbital period of somewhat less than
2 hours derived from spectroscopy. Kato (2009) reported a 
superhump period of 107.8 min. 

\begin{figure}
\centerline{\hbox{\psfig{figure=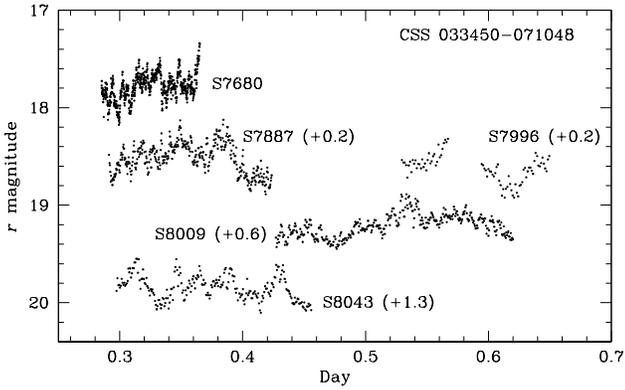,width=8.4cm}}}
  \caption{Individual light curves of CSS0334-07.  The light curve
of run S7680 is displayed at the correct brightness; vertical offsets
for each light curve are given in brackets.}
 \label{css0334fig1}
\end{figure}

Our observations
are listed in Tab.~\ref{survey7tab1}; the light curves seen in Fig.~\ref{css0334fig1}
possess typical flickering with a range of $\sim 0.5$ mag, but no modulation on
time scales of the proposed orbital period.

\subsection{CSS0345-01 (CSS090219:034515-015216)}

CSS0345-01 was first reported in the CSS survey in 2009 February as
CSS090219:034515-015216. It has a quiescent magnitude around $V \sim 19$
and has had seven recorded outbursts since 2005, reaching up to $V \sim 15.7$.

\begin{figure}
\centerline{\hbox{\psfig{figure=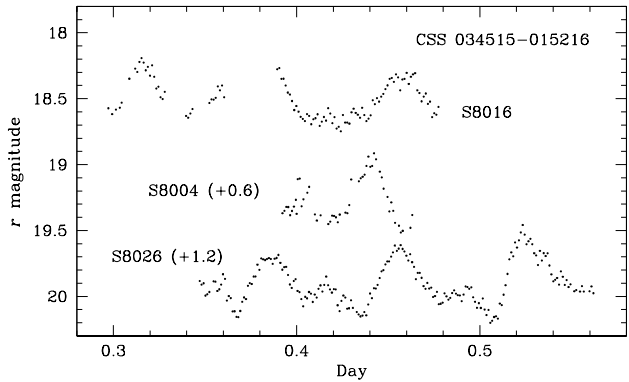,width=8.4cm}}}
  \caption{Individual light curves of CSS0345-01. The light curve
of run S8016 is displayed at the correct brightness; vertical
offsets for each light curve are given in brackets. }
 \label{css0345fig1}
\end{figure}

\begin{figure}
\centerline{\hbox{\psfig{figure=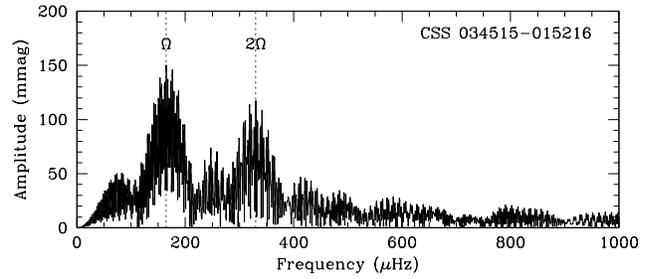,width=8.4cm}}}
  \caption{Fourier transform of CSS0345-01 during 2010 November. The orbital 
frequency ($\Omega$) and its first harmonic ($2 \Omega$) are labelled and marked
by vertical dashed lines.}
 \label{css0345fig2}
\end{figure}

\begin{figure}
\centerline{\hbox{\psfig{figure=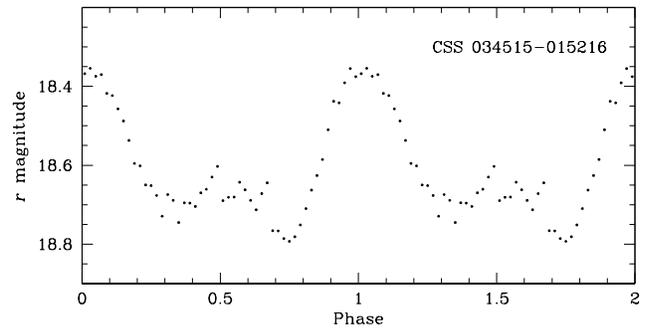,width=8.4cm}}}
  \caption{The average light curve of CSS0345-01, folded on the 
ephemeris given in Eq.~\ref{ephcss0345}.}
 \label{css0345fig3}
\end{figure}

We observed CSS0345-01 during quiescence in 2010 November (see Tab.~\ref{survey7tab1})
when it was around $r \sim 18.6$. The light curves are shown in Fig.~\ref{css0345fig1}
and resemble those of SSS0221-26, characteristic of a low mass transfer dwarf nova.
The FT of the combined observing runs (Fig.~\ref{css0345fig2})
reveals two distinct
peaks at $164.81 \pm 0.04$ $\mu$Hz and $330.01 \pm 0.05$ $\mu$Hz, which we assign as the 
fundamental and first harmonic, respectively, of the orbital frequency.
The average light curve of CSS0345-01, folded on the orbital period
of 101.1 min, is shown in Fig.~\ref{css0345fig3}. The emphemeris for maximum
light is:

\begin{equation}
    {\rm HJD_{max}} = 245\,5527.31408 + 0.07018\,(4)\, {\rm E.}
\label{ephcss0345}
\end{equation}

\subsection{SSS0617-36 (SSS100511:061754-362655)}

SSS100511:061754-362655 (SSS0617-36 hereafter) was first reported
by the SSS survey in 2010 May. It has regular outbursts reaching 
outburst amplitudes of up to 4 magnitude, varying from $V \sim 18$ to
$\sim$14.2.

\begin{figure}
\centerline{\hbox{\psfig{figure=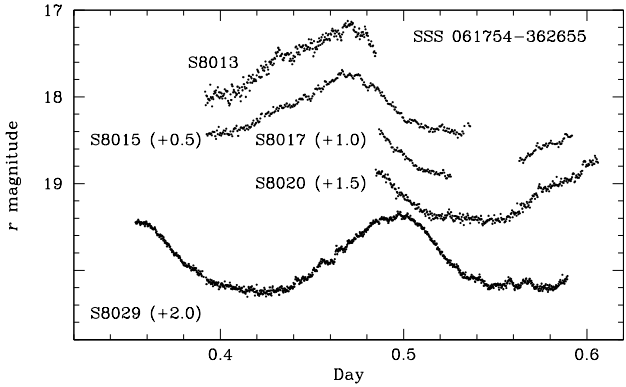,width=8.4cm}}}
  \caption{Individual light curves of SSS0617-36.  The light curve
of run S8013 is displayed at the correct brightness; vertical offsets
of each light curve are given in brackets.}
 \label{sss0617fig1}
\end{figure}

\begin{figure}
\centerline{\hbox{\psfig{figure=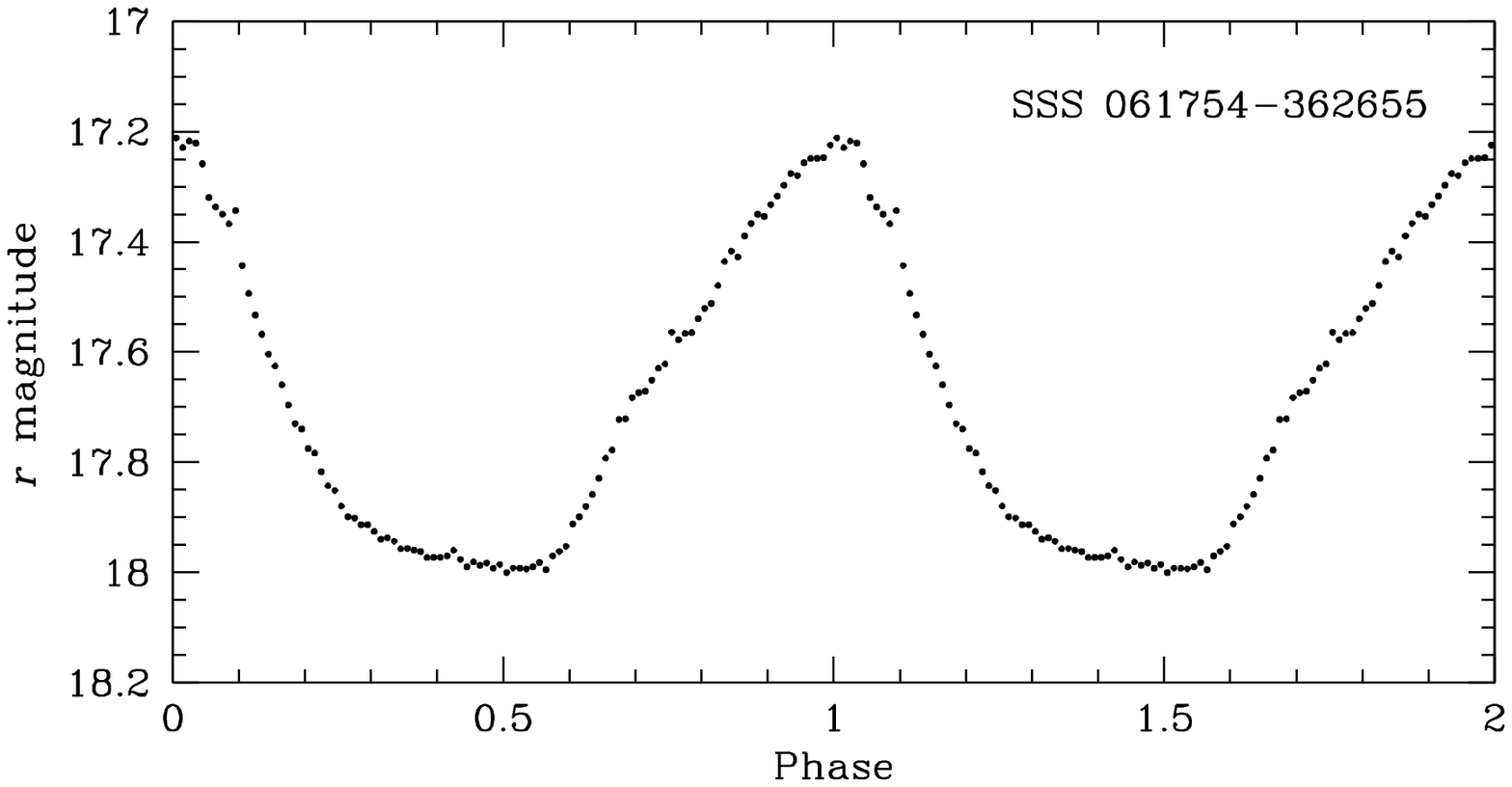,width=8.4cm}}}
  \caption{The average light curve of SSS0617-36, folded on the 
ephemeris given in Eq.~\ref{ephsss0617}.}
 \label{sss0617fig2}
\end{figure}

We observed SSS0617-36 during quiescence in 2010 November when the system
averaged around $r \sim 17.7$. The light curves are shown in Fig.~\ref{sss0617fig1}.
The light curves are characterised by a large-amplitude orbital modulation (by about a factor
of 2 in brightness) and low-amplitude flickering. The orbital modulation, clearly shown
in the average light curve (Fig.~\ref{sss0617fig2}), is non-sinusoidal with
power in the Fourier transform at the fundamental frequency (80.72 $\mu$Hz) 
and the first two harmonics (161.53 $\mu$Hz and 242.20 $\mu$Hz, respectively).
The light curve is unusual for a CV and at this stage we cannot provide a
specific class.
The orbital emphemeris for maximum light is

\begin{equation}
    {\rm HJD_{max}} = 245\,5525.46712 + 0.14335\,(5)\, {\rm E.}
\label{ephsss0617}
\end{equation}

\subsection{CSS0810+00 (CSS100108:081031+002429)}

\begin{figure}
\centerline{\hbox{\psfig{figure=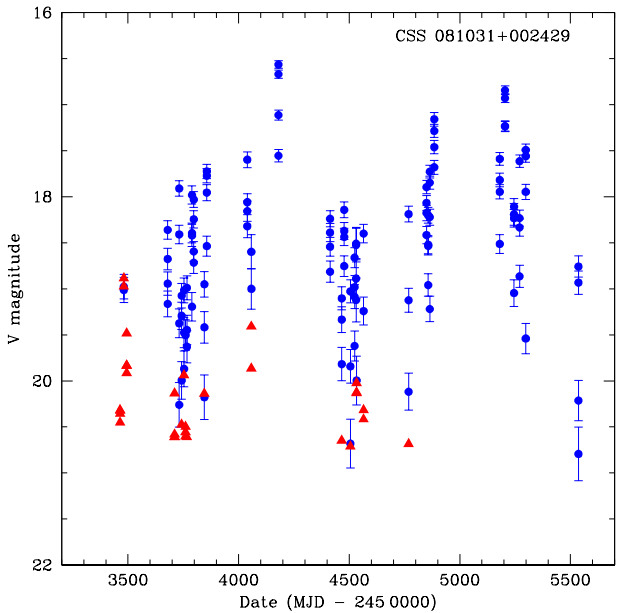,width=8.4cm}}}
  \caption{The long term CSS light curve of CSS0810+00. The filled triangles represent
upper limits; times are given in heliocentric modified Julian date.}
 \label{css0810fig2}
\end{figure}

\begin{figure}
\centerline{\hbox{\psfig{figure=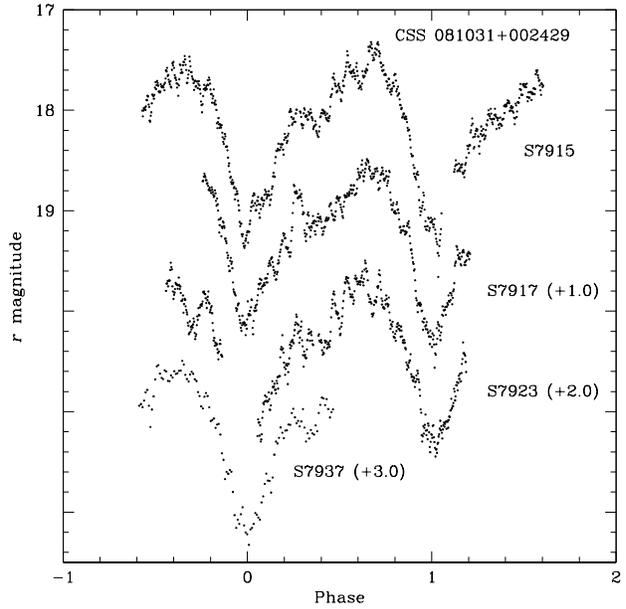,width=8.4cm}}}
  \caption{Individual light curves of CSS0810+00, folded on
the ephemeris given in Eq.~\ref{ephcss0810}.  The light curve
of run S7915 is displayed at the correct brightness; vertical
offsets for each light curve are given in brackets.}
 \label{css0810fig1}
\end{figure}

CSS0810+00 is CSS100108:081031+002429; its 
long-term light curve shows brightness variations 
from $V \sim 21$ to $V \sim 17.3$ on a range of time 
scales (Fig.~\ref{css0810fig2}, data kindly provided by Dr.~Andrew 
Drake). From that light curve alone it would 
be difficult to state what kind of CV this is. Our observations are 
listed in Tab.~\ref{survey7tab1} and the light curves are given 
in Fig.~\ref{css0810fig1} and were obtained during a high
state of CSS0810+00. There are deep minima with 
depths $\sim$ 1.5 mag and irregular profiles recurring with 
period 116.15 min. The light curve is very similar to that
of the polar V834 Cen in the V band which also has $\sim 2$ mag 
range around the orbit (Sambruno et al.~1991). This is compatible 
with the long term light curve shown in Fig.~\ref{css0810fig2}. 
The ephemeris for minimum light is 

\begin{equation}
    {\rm HJD_{min}} = 245\,5274.2967 + 0.080660\,(2)\, {\rm E.}
\label{ephcss0810}
\end{equation}

\subsection{CSS0826-00 (CSS080306:082655-000733)}

\begin{figure}
\centerline{\hbox{\psfig{figure=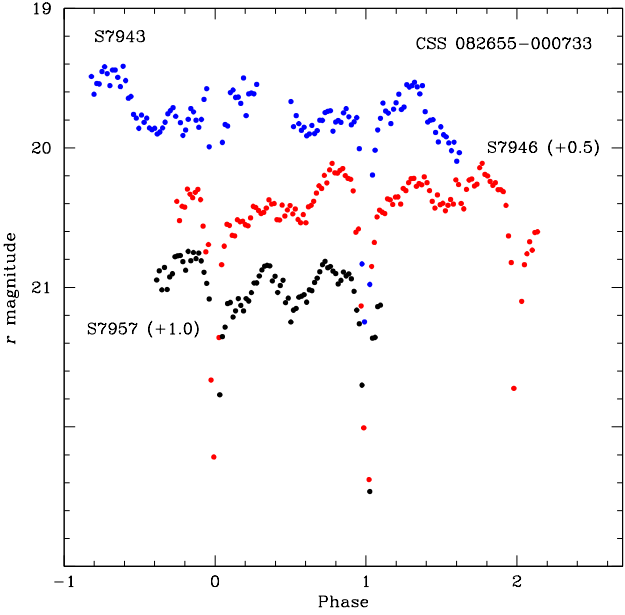,width=8.4cm}}}
  \caption{Individual light curves of CSS0826-00, folded on
the ephemeris given in Eq.~\ref{ephcss0826}.  The light curve of run S7943
is displayed at the correct brightness; vertical offsets for each light curve
are given in brackets.}
 \label{css0826fig1}
\end{figure}

CSS0826-00 is CSS080306:082655-000733 and has had two 
observed outbursts raising it from quiescent magnitude 
$r \sim 20$ to 16.3 and 17.7 respectively. Our observations 
were made at quiescence with $r = 19.9$ and are listed in 
Tab.~\ref{survey7tab1}. The light curves, displayed in 
Fig.~\ref{css0826fig1}, show that CSS0826-00 is an eclipsing 
system, descending to $r \sim 21.0$ at mid-eclipse with $P_{orb} = 86.05$ 
min. The eclipse ephemeris is 

\begin{equation}
    {\rm HJD_{min}} = 245\,5293.3415 + 0.05976\,(1)\, {\rm E.}
\label{ephcss0826}
\end{equation}

  Signs of a superhump drifting through the light curve can be 
seen in Fig.~\ref{css0826fig1}, and the FT confirms this, giving 
a negative superhump period of 83.63 min; the superhump ephemeris is 

\begin{equation}
    {\rm HJD_{SH, max}} = 245\,5293.2379 + 0.05807\,(3)\, {\rm E.}
\label{ephcss0826sh}
\end{equation}

\begin{figure}
\centerline{\hbox{\psfig{figure=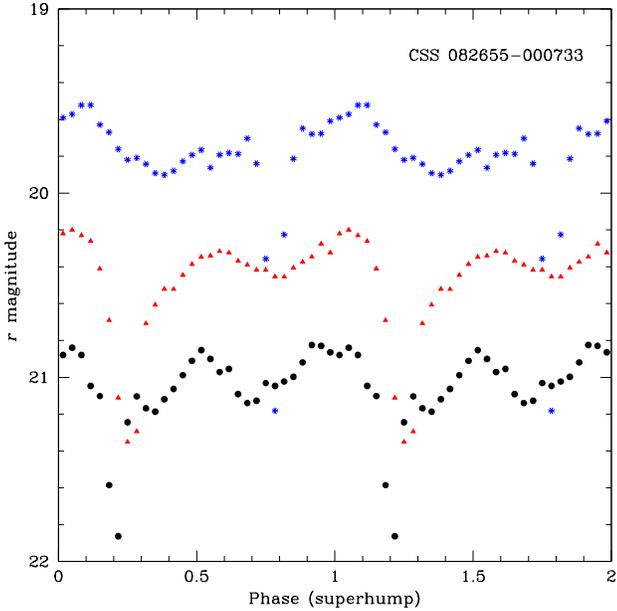,width=8.4cm}}}
  \caption{Nightly-averaged light curves of CSS0826-00, folded on
the negative superhump ephemeris given in Eq.~\ref{ephcss0826sh}.  The light 
curve of run S7943 is displayed at the correct brightness. Runs S7946 and S7957 have been 
displaced vertically by 0.5 and 1.0 mag, respectively.}
 \label{css0826fig2}
\end{figure}

The nightly-averaged light curves folded on the superhump ephemeris are shown in 
Fig.~\ref{css0826fig2}.  Thus CSS 0826-00 is another SU UMa type dwarf nova with 
negative superhumps persisting in quiescence, with a negative superhump deficiency
of $\epsilon^- = -0.0279$.

\subsection{CSS1028-08 (CSS090331:102843-081927)}

\begin{figure}
\centerline{\hbox{\psfig{figure=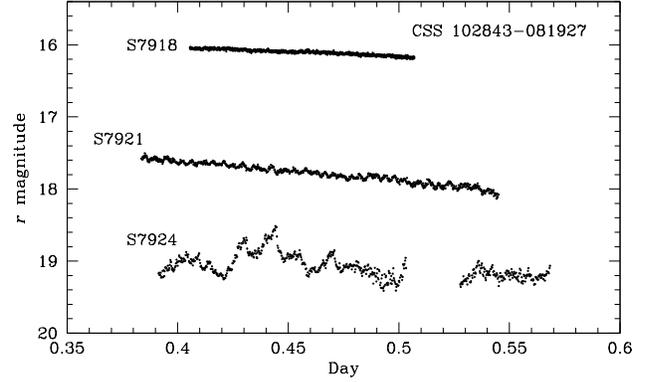,width=8.4cm}}}
  \caption{Individual light curves of CSS1028-08.  All the light curves
are displayed at their correct brightness.}
 \label{css1028fig1}
\end{figure}

CSS1028-08 is CSS 090331:102843-081927 which in the CSS 
light curve has outbursts at intervals as short as 300 d, but 
with frequent short-lived dips to $V > 20.5$. Our observations are 
listed in Tab.~\ref{survey7tab1} and the light curves are shown in 
Fig.~\ref{css1028fig1}. CSS1028-08 was descending to quiesence from
 an outburst (which was missed in the Catalina survey). We do not 
see any excursions to faint magnitudes in quiescence, and are unable 
to ascertain the cause of the ones in the CSS light curve and suspect that
because the star has a close companion the CSS photometry has difficulty in
resolving the two. In our photometry the two stars are completely resolved.

   CSS1028-08 was observed during superoutburst by Kato et al (2009), 
who named it OT J1028 and found a superhump period of 54.85 min, which 
puts it in the rare class of ultra-short period CVs, similar to V485 Cen 
and EI Psc, probably with a lowered H/He abundance ratios. The absence of
superhumps in run S7918 suggests we observed CSS1028-08 during a normal
outburst.

\begin{figure}
\centerline{\hbox{\psfig{figure=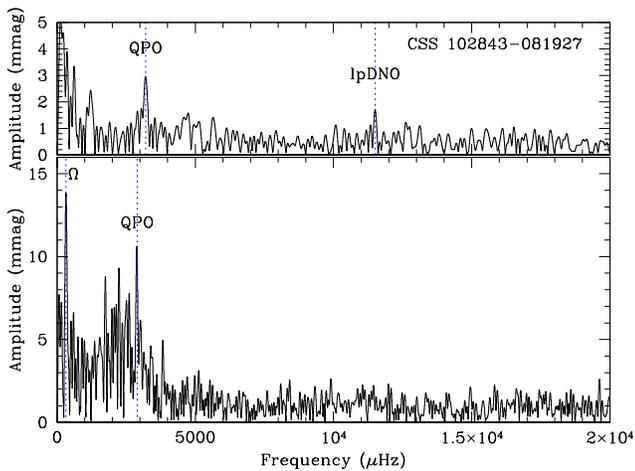,width=8.4cm}}}
  \caption{Fourier transforms of CSS1028-08; run S7918 is displayed in the upper panel, 
run S7921 is shown in the lower panel. Quasi-periodic and longer-period dwarf nova 
oscillations are marked and labelled. The orbital modulation ($\Omega$) is marked in the lower
panel.}
 \label{css1028fig2}
\end{figure}

\begin{figure}
\centerline{\hbox{\psfig{figure=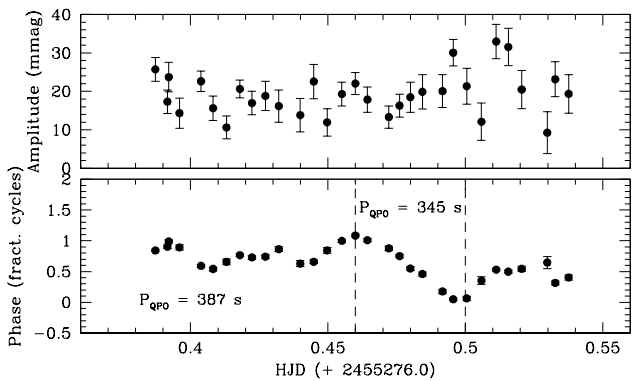,width=8.4cm}}}
  \caption{The phase - amplitude diagram for run S7921 of CSS1028-08, relative to the 
QPO period of 387 s. Between the vertical dashed lines the phase shift indicates the period
had changed to 345 s.}
 \label{css1028fig3}
\end{figure}

   The FTs of runs S7918 and S7921 are shown in Fig.~\ref{css1028fig2}.  The highest
peak in the FT of run S7921 is at 52.1 $\pm$ 0.6 min, which we identify as the orbital
period; this gives a positive superhump excess of $\epsilon^+ = 0.0530$. 
The FT of run S7918 shows two short-period modulations which are at 312 s and 86.9 s,
which we interpret respectively as a quasi-periodic oscillation (QPO) and a longer-period dwarf
nova oscillation of the kinds frequently seen in outbursting dwarf novae (Warner 2004).
A similar peak at 345 s is seen in the FT of run S7921, with a lot of additional power at 
nearby frequencies. The reason for the complexity of the FT is seen in the phase - amplitude
diagram shown in Fig.~\ref{css1028fig3}; it is clear that there is a persistent oscillation which
both drifts and jumps in frequency, again typical of such rapid oscillations. Of particular interest
is the phase change seen in Fig.~\ref{css1028fig3} between HJD 245\,5276.45 and 245\,5276.50 where
the QPO period changed from 387 s to 345 s; their frequency difference 
is approximately equal to the orbital frequency and suggests a 
temporary change from a sidereal to a synodic (reprocessed) QPO signal.

\subsection{CSS1300+11 (CSS080702:130030+115101)}

\begin{figure}
\centerline{\hbox{\psfig{figure=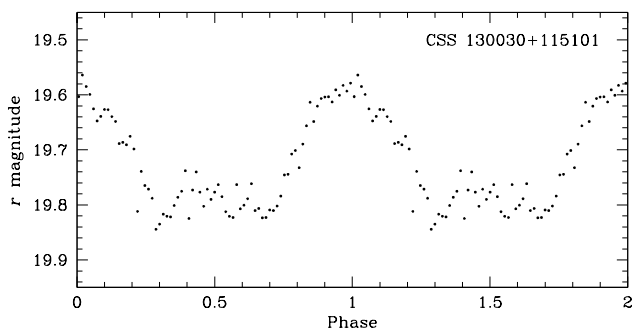,width=8.4cm}}}
  \caption{The average light curve of CSS1300+11, folded on the 
ephemeris given in Eq.~\ref{ephcss1300}.}
 \label{css1300fig1}
\end{figure}

CSS1300+11, is CSS080702:130030+115101 and shows a 
superoutburst in the CSS long-term light curve, 
reaching $V = 13.9$ in 2008 July from a quiescent level of $V \sim 20$. 
It was observed by Kato et al (2009) who measured a superhump 
period of 92.72 min.  We observed it in 2010 June to find the 
orbital period. Our observations are listed in Tab.~\ref{survey7tab1} and 
the average light curve is shown in Fig.~\ref{css1300fig1}. The 
orbital period is 90.24 min, which gives a fractional superhump 
period excess of $\epsilon^+ = 0.0274$; this is in agreement with the 
general correlation between $\epsilon^+$ and $P_{orb}$ 
(Patterson et al.~2003). The orbital ephemeris for the time 
of maximum brightness is

\begin{equation}
    {\rm HJD_{max}} = 245\,5338.3346 + 0.06267\,(1)\, {\rm E.}
\label{ephcss1300}
\end{equation}

\subsection{HV Vir (CSS080227:132103+015329)}

CSS1321+01 = HV Vir, is CSS 080227:132103+015329, had 
a superoutburst in 2002 January after a decade of quiescence, reaching 
$V = 14.15$ in the CSS light curve, from a general quiescent level of 
$V \sim 19.2$. This star has been well observed during supermaxima, so its 
superhump period is well determined at 83.82 min, but the observations 
at quiesence by Patterson et al.~(2003) showed two periodicities in the FT: 
a double humped definite one at $P = 82.18$ min which is $P_{orb}$ and another, 
less certain at 128.6 min or its alias at 117.9 min.

\begin{figure}
\centerline{\hbox{\psfig{figure=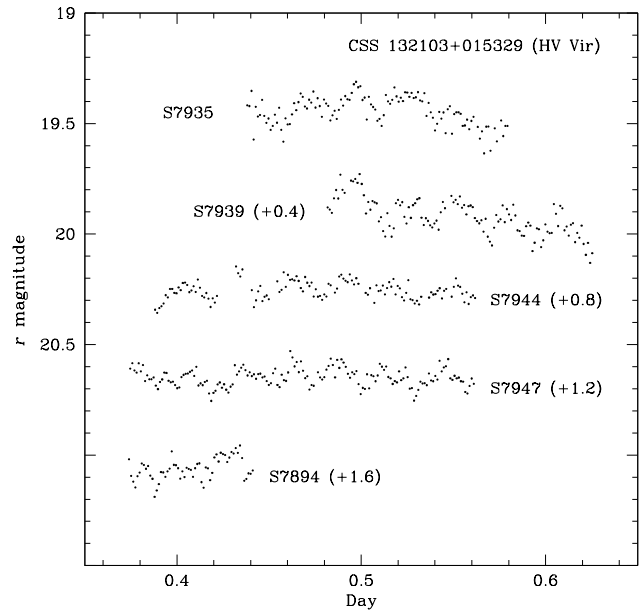,width=8.4cm}}}
  \caption{Individual light curves of CSS1321+01.  The light curve
of run S7935 is displayed at the correct brightness; vertical
offsets for each light curves are given in brackets.}
 \label{css1321fig1}
\end{figure}

\begin{figure}
\centerline{\hbox{\psfig{figure=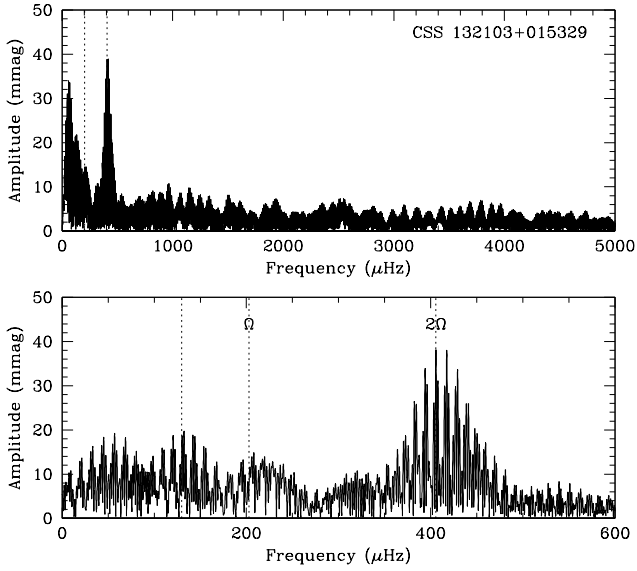,width=8.4cm}}}
  \caption{Fourier transform of CSS1321+01 during 2010 April. The orbital
frequency ($\Omega$) and its first harmonic ($2 \Omega$) are labelled and
marked by vertical dashed lines. The 127-min (or 117-min) modulation of unknown 
origin is also marked by a vertical dashed line.}
 \label{css1321fig2}
\end{figure}

We observed HV Vir in quiescence in 2010 April; our observations are listed 
in Tab.~\ref{survey7tab1}
and the light curves are shown in Fig.~\ref{css1321fig1}. The FT of these light 
curves, shown in Fig.~\ref{css1321fig2}, contains a strong signal at the first 
harmonic of the orbital frequency (i.e. at $2 \Omega$) 
and a detectable signal at $\Omega$ itself; but in addition there is a window pattern 
at lower frequency, stronger than the $\Omega$ modulation, with period 126.0 min 
or 117.4 min. Within the uncertainties these are identical to the signal and 
its alias observed by Pattersion et al. These independent data sets confirm 
the existence of a luminosity modulation at $\sim 127$ min or its alias at 117 min, the 
origin of which is not known.

\subsection{CSS1404-10 (CSS080623:140454-102702)}

\begin{figure}
\centerline{\hbox{\psfig{figure=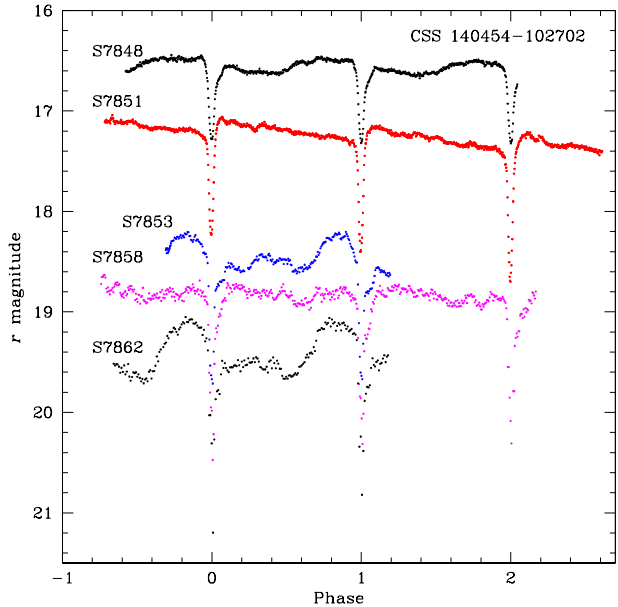,width=8.4cm}}}
  \caption{Individual light curves of CSS1404-10, folded on
the ephemeris given in Eq.~\ref{ephcss1404}.  All light curves
are displayed at their correct brightness.}
 \label{css1404fig1}
\end{figure}

CSS1404-10, CSS080623:140454-102702) has a quiescent 
brightness level of $V \sim 19.5$ in the CSS light curve, 
with an outburst to $V \sim 15.0$ and others to $V \sim 16.5$
at intervals of  $\sim 500$ d. Our observations were made at 
quiescence in 2009 February and June, and during the decline of a 
superoutburst in 2009 March: they are listed in Tab.~\ref{survey7tab1}. 
The light curves are displayed in Fig.~\ref{css1404fig1} and show 
narrow eclipses more than one magnitude deep in quiescence. 
The orbital period is 85.794 min and the eclipse ephemeris is 

\begin{equation}
    {\rm HJD_{min}} = 245\,4911.49044 + 0.0595790\,(2)\, {\rm E.}
\label{ephcss1404}
\end{equation}

\begin{figure}
\centerline{\hbox{\psfig{figure=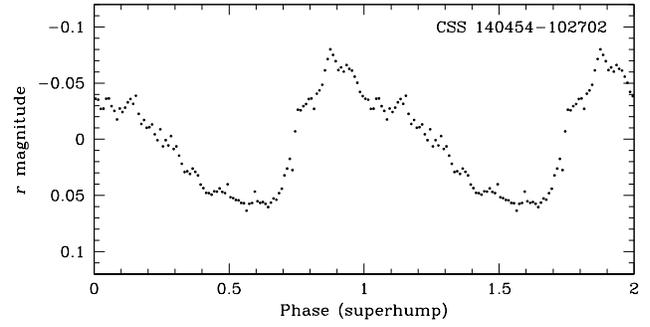,width=8.4cm}}}
  \caption{The average light curve of CSS1404 during superoutburst
(runs S7848, S7851; eclipses removed), folded on the superhump
period of 0.06098 d.}
 \label{css1404fig2}
\end{figure}

Superhumps were evident during the superoutburst. Their period is 
87.81 min, with a resultant superhump period excess of 
$\epsilon^+ = 0.0238$. The average superhump light curve, folded on 
the superhump period and with the orbital eclipses excised, showing the 
typical saw-tooth shape, is given in Fig.~\ref{css1404fig2}.

\subsection{CSS1443-17 (CSS090418:144342-175550)}

\begin{figure}
\centerline{\hbox{\psfig{figure=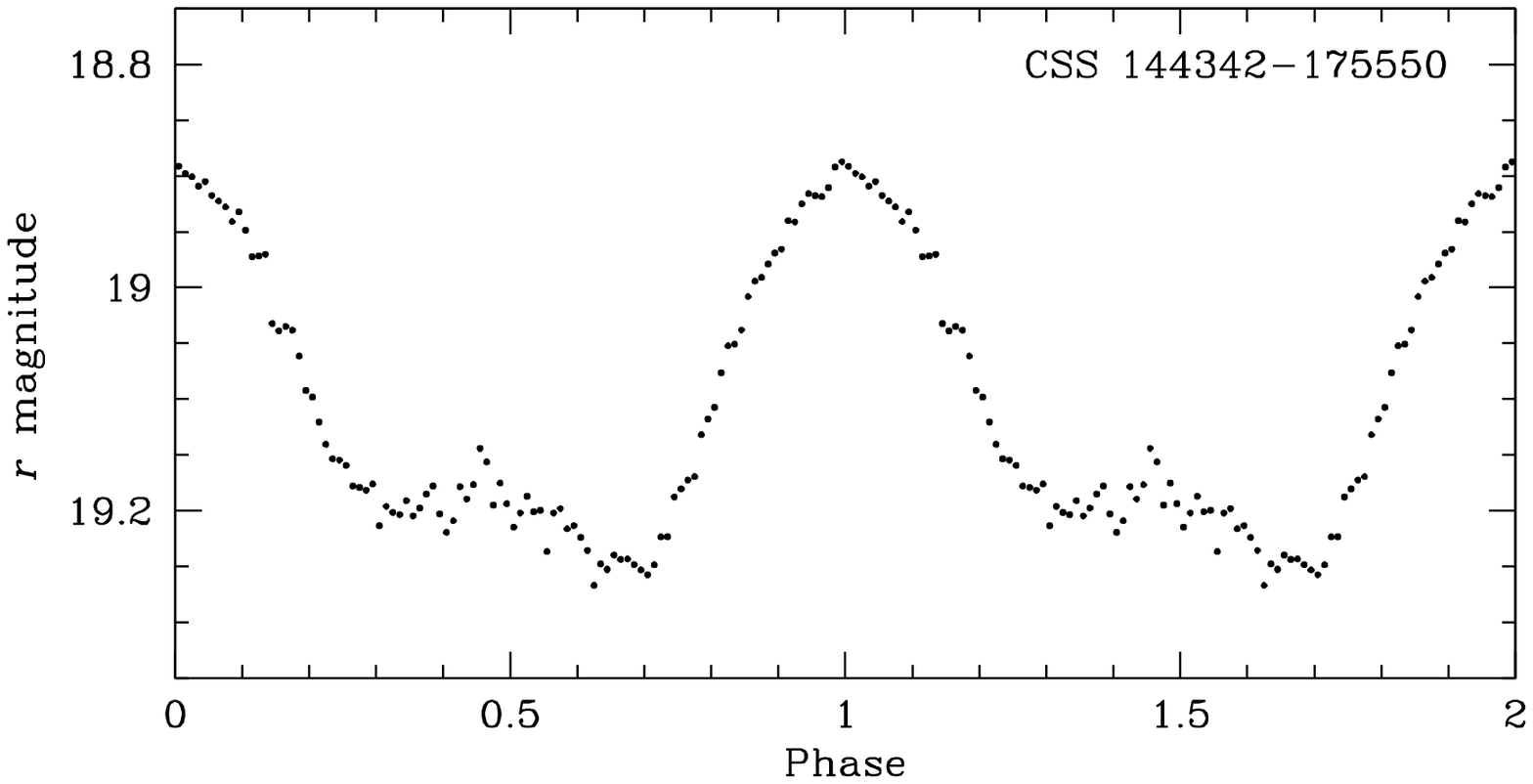,width=8.4cm}}}
  \caption{The average light curve of CSS1443-17, folded on the 
ephemeris given in Eq.~\ref{ephcss1443}.}
 \label{css1443fig1}
\end{figure}

CSS1443-17, or CSS 090418:144342-175550, has one superoutburst 
in the CSS light curve, in 2009 April, which was observed by Kato 
et al (2009), where it is listed as OT J1443, and who found 
superhumps with a period of 103.77 min. We observed it at quiescence 
in 2009 June - the observations are listed in Tab.~\ref{survey7tab1} - and 
found it to have a strong orbital modulation with a range of 0.3 mag, 
as seen in the average light curve in Fig.~\ref{css1443fig1}. 
The orbital period is 101.1 min and the ephemeris for maximum light is 

\begin{equation}
    {\rm HJD_{max}} = 245\,5000.2092 + 0.07019\,(17)\, {\rm E.}
\label{ephcss1443}
\end{equation}

The superhump excess is $\epsilon^+ = 0.0264$.

\subsection{CSS1503-22 (CSS100216:150354-220711)}

\begin{figure}
\centerline{\hbox{\psfig{figure=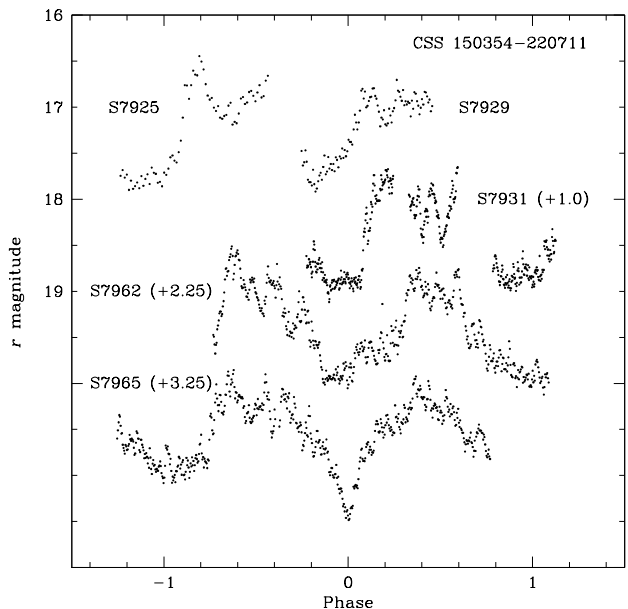,width=8.4cm}}}
  \caption{Individual light curves of CSS1503+22, folded on
the ephemeris given in Eq.~\ref{ephcss1503}.  The light curves
of run S7925 and S7929 are displayed at the correct brightness. Runs S7931,
S7962 and S7965 have been displaced vertically by 1, 2.25 and 3.25 mag,
respectively.}
 \label{css1503fig1}
\end{figure}

CSS1503-22, known as CSS100216:150354-220711, has a CSS light curve 
in which the star was at an average brightness $V \sim 19.9$ for about 5.5 years 
and then went into a high state with large excursions at an average $V \sim 17.5$. 
Our observations were made during the high state and are listed in Tab.~\ref{survey7tab1}. 
The light curves are shown in Fig.~\ref{css1503fig1}, phased on the 133.38 min 
period that we found in the FT. The existence of distinct window patterns at the
first and third harmonics enables a unique choice among the aliases of the fundamental
frequency. From its general optical behaviour we suspected 
this star to be a polar. We communicated the discovery to Julian Osborne at Leicester 
University and he confirmed from Swift X-Ray observations that it is indeed 
a magnetically accreting CV. Our ephemeris for the times of minimum light is

\begin{equation}
    {\rm HJD_{min}} = 245\,5300.5757 + 0.092622\,(15)\, {\rm E.}
\label{ephcss1503}
\end{equation}

\subsection{CSS1528+03 (CSS090419:152858+034912)}

\begin{figure}
\centerline{\hbox{\psfig{figure=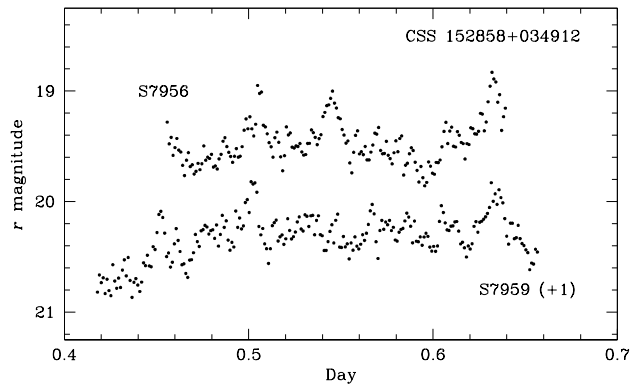,width=8.4cm}}}
  \caption{Individual light curves of CSS1528+03.  The light curve
of run S7956 is displayed at the correct brightness. Run S7959
has been displaced vertically by 1 mag for display purposes.}
 \label{css1528fig1}
\end{figure}

CSS1528+03, or CSS 090419:152858+034912 in the CSS catalogue, is also 
in the Sloan survey second release (Szkody et al.~2003) as SDSS J1528+0349, 
where its spectrum shows it be a dwarf nova. Its CSS light curve shows 
variations from $V$ = 18.7 to $\sim 21$ with rare outbursts to 17th magnitude. 
Our photometric observations, the first of which was taken during outburst, are 
listed in Tab.~\ref{survey7tab1}, and the two longest runs (both at quiescence)
are shown as light curves in Fig.~\ref{css1528fig1}. There is 
strong flickering but no obvious modulation within our longest 
($\sim 6$ h) run. Either this CV has a very long orbital period or it is of low inclination.

\subsection{CSS1626-12 (CSS090419:162620-125557)}

\begin{figure}
\centerline{\hbox{\psfig{figure=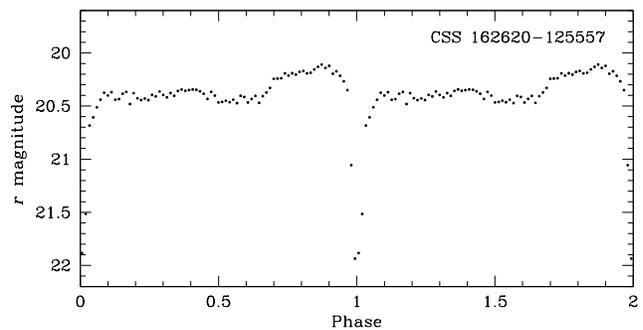,width=8.4cm}}}
  \caption{The average light curve of CSS1626-12, folded on the 
ephemeris given in Eq.~\ref{ephcss1626}.}
 \label{css1626fig1}
\end{figure}

CSS1626-12 is CSS090419:162620-125557 and has a CSS light 
curve which is undersampled but shows outbursts to $V \sim 17$ 
on time scales $\sim 300$ d. It is not detectable by CSS at 
quiesence, reaching $V > 20.4$. Our observations are listed 
in Tab.~\ref{survey7tab1} and show that its un-eclipsed magnitude 
is in fact at $r = 20.4$. The average light curve is given in 
Fig.~\ref{css1626fig1}, where narrow eclipses to $r \sim 22.0$ 
are seen, recurring on the period $P_{orb} = 108.7$ min, with an ephemeris

\begin{equation}
    {\rm HJD_{min}} = 245\,5352.4721 + 0.07546\,(2)\, {\rm E.}
\label{ephcss1626}
\end{equation}

The light curve has a low amplitude orbital hump peaking at 
orbital phase 0.8 before being cut into by the eclipse.

\subsection{EG Aqr (CSS080101:232519-081819)}

\begin{figure}
\centerline{\hbox{\psfig{figure=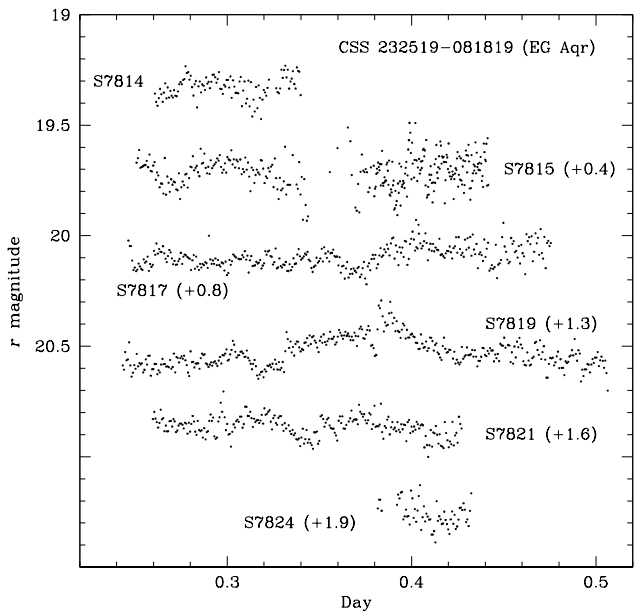,width=8.4cm}}}
  \caption{Individual light curves of CSS2325-08.  The light curve
of run S7814 is displayed at the correct brightness; vertical offsets
for each light curve are given in brackets.}
 \label{css2325fig1}
\end{figure}

\begin{figure}
\centerline{\hbox{\psfig{figure=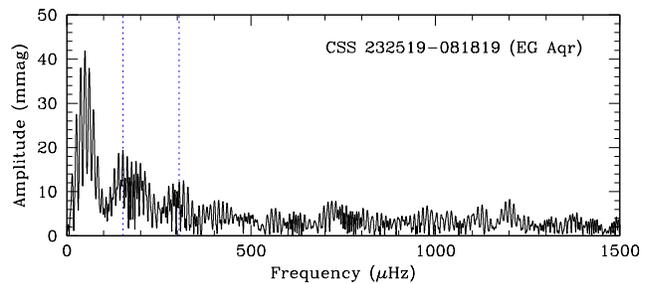,width=8.4cm}}}
  \caption{Fourier transform of the combined runs S7817, S7819 and S7821 of CSS2325-08.}
 \label{css2325fig2}
\end{figure}

CSS2325-08, or CSS 080101:232519-081819, is one of the first 
CVs to be found in the CSS survey and was a rediscovery of the 
known dwarf nova EG Aqr. Imada et al (2008) observed a superoutburst 
in EG Aqr in 2006 November and found a superhump period of 113.51 min. 
We observed EG Aqr at a quiescence brightness of $r \sim 19.3$ in 
2008 October; the details are given in Tab.~\ref{survey7tab1} and 
the light curves in Fig.~\ref{css2325fig1}. The FT of the 
combination of the best runs, S7817, S7819 and S7821, from consecutive 
nights is shown in Fig.~\ref{css2325fig2}.

In the FT there is only one pair of frequencies that are at a 2:1 ratio,
which gives a period of 109.4 min and its first harmonic, which are indicated
in Fig.~\ref{css2325fig2} by the dashed vertical lines. We identify this with
$P_{orb}$ which gives a $\epsilon^+ = 0.0378$. The ephemeris for maximum
light is:

\begin{equation}
    {\rm HJD_{max}} = 245\,4757.2445 + 0.07596\,(8)\, {\rm E.}
\label{ephcss2325}
\end{equation}

\section{Discussion}

\begin{figure*}
\centerline{\hbox{\psfig{figure=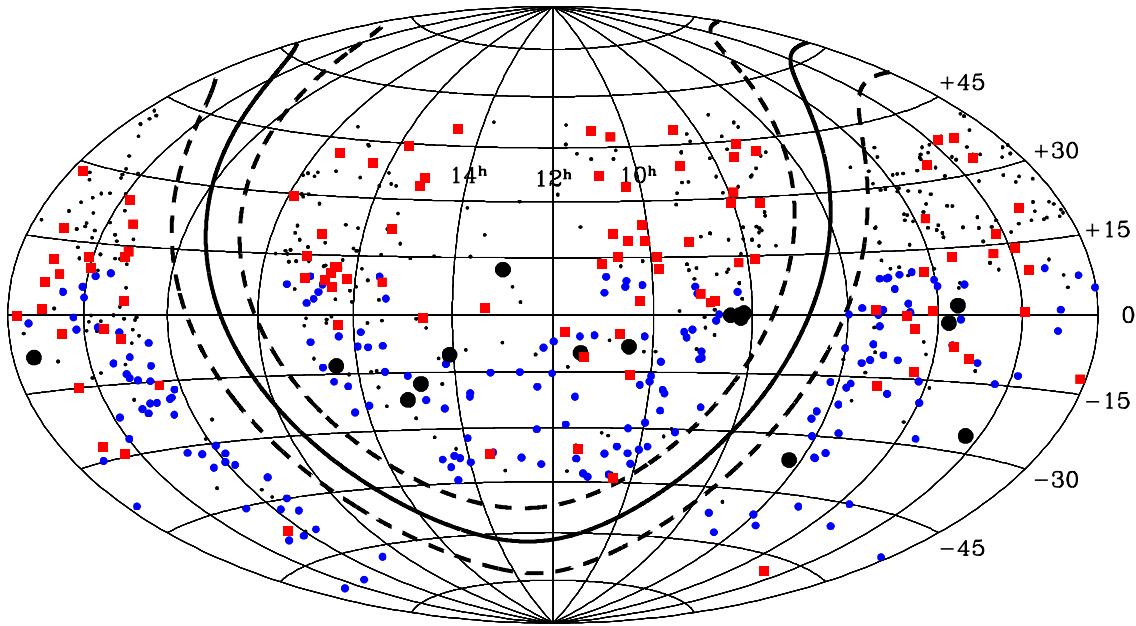,width=16cm}}}
  \caption{The distribution in an equatorial aitoff projection
of CVs identified in the Catalina Real-Time Transient Survey. The Galactic Plane
($b = 0^{\circ}$) is marked by the solid line, the Galactic latitudes of $b = \pm 10^{\circ}$
are shown as dotted lines. The blue filled circles (colour is available only for the
online version) mark those CVs observable from the SAAO with the 40-inch and
74-inch reflector and the UCT CCD (Declination (2000.0) $< +10^{\circ}$).
Filled red triangles and squares display CVs with periods known prior to our 
photometric follow-up ($N = 100$), whereas the large black filled circles
mark CVs with periods derived following the UCT CCD CV survey ($N = 15$). }
 \label{cssdistro}
\end{figure*}

Our observations have assessed the properties of a further 20 faint CVs, 
resulting in the direct measurement of 15 orbital periods; the measurement 
of 2 further positive superhump periods with consequent determination of 
3 fractional superhump excesses; and the measurement of two negative superhump 
periods. Among these are CSS 0322+02 which has both positive and negative 
superhumps in quiescence, and CSS 0826-00, an eclipsing system with negative 
superhumps in quiescence, which potentially gives the possibility of 
deducing the brightness distribution of the negative superhump source.

   CSS 0810+00 and CSS 1503-22 are new polars. SDSS 0919+08 has a known 
260 s brightness modulation that could be interpreted either as non-radial 
pulsation or rotation; our discovery of an occasional additional 214 s 
periodicity points to the former explanation. CSS 0802+01 (HV Vir) had 
a previously suspected periodic signal near 120 min; we have observed 
a similar modulation, the origin of which is not known.
   A summary of the newly measured periods is included in Tab.~\ref{survey7tab3}.

\begin{table*}
 \centering
  \caption{Summary of results}
  \begin{tabular}{@{}lllllcl@{}}
 Object      & Type    & $P_{orb}^{\ast}$  & $P_{{SH}^+}^{\dag}$  & $P_{{SH}^-}$ & {\em r} & Remarks \\  
             &         & (h)       & (h)     & (h)      &           & \\[10pt]
SDSS0805+07 & CV      & 5.489\,(12)    &         &          &     17.9  &  \\
SDSS0904+03 & CV      & 1.433358\,(1)  &         &          &     19.2  & Eclipsing \\
SSS0221-26  & DN      & 1.692\,(2)     &         &          &     19.3: & Possible alias at 1.818 h. \\
CSS0332+02  & DN SU   & 1.469\,(1)     & 1.5030\,(2) & 1.4387\,(4) &   20.2  & Superhumps in quiescence\\
CSS0345-01  & DN      & 1.684\,(1)     &             &          &     18.6: &  \\
SSS0617-36  & CV      & 3.4404\,(12)   &             &          &     17.7  & \\
CSS0810+00  & Polar   & 1.9358\,(1)    &             &          &     18.2  & High state \\
CSS0826-00  & DN SU   & 1.4342\,(2)    &             & 1.394\,(1) &     20.0  & Eclipsing, superhumps in quiescence \\
CSS1028-08  & DN      & 0.868\,(10)    & [0.914]     &          & 16.1-19.0 & $P_{{SH}^+}$ from Kato et al. (2009)  \\
CSS1300+11  & DN SU   & 1.5041\,(2)    & [1.545]     &          &     19.8: & $P_{{SH}^+}$ from Kato et al. (2009) \\
CSS1404-10  & DN SU   & 1.42990\,(1)   & 1.464\,(1)  &          & 16.6-19.6 & Eclipsing, superhumps in outburst \\
CSS1443-17  & DN      & 1.685\,(4)     & [1.7295]    &          &     19.1  & $P_{{SH}^+}$ from Kato et al. (2009) \\
CSS1503-22  & Polar   & 2.2229\,(4)    &             &          &     17.2  & High state   \\
CSS1626-12  & DN      & 1.811\,(1)     &             &          &     20.4  & Eclipsing \\
CSS2325-08  & DN SU   & 1.823\,(2)     & [1.892]     &          &     19.3  & $P_{{SH}^+}$ from Imada et al. (2008)  \\
\end{tabular}
\label{survey7tab3}
{\footnotesize 
\newline 
$^{\ast}$ Uncertainties are given between brackets for the last significant decimal(s). $^{\dag}$ Literature
values are shown between square brackets.\hfill}
\end{table*}

Population syntheses combined with evolutionary computations of 
binary stars passing through the common envelope stage have long 
predicted that there should be a maximum space density of CVs near 
the minimum orbital period at $P_{orb} \sim 80$ min, which is where 
$dP_{orb}/dt$ = 0 (e.g. Kolb \& Baraffe 1999). But until recently this 
``period spike'' had not been confirmed observationally. However, 
G\"ansicke et al.~(2009), from large numbers of faint CVs newly discovered 
in the SDSS survey, at last found evidence for the spike -- because it occurs 
at small $P_{orb}$ where the mass transfer rate is low the CVs have 
low accretion luminosity (typically $M_V \sim 11.5$), it is necessary 
to reach fainter than $V \sim 18 $ before the predicted population can be sampled. 
This was verified in another way by Wils et al.~(2009) who cross-correlated 
a number of photometric and astrometric catalogues, finding 64 new CV candidates 
which, from the outburst amplitudes and frequencies of these and CRTS objects, 
imply that the CVs of faint apparent magnitudes are not simply more 
distant -- there must also be a population of intrinsically faint systems. 
In affirmation, the CVs found in the Hamburg Quasar Survey, which have 
$B < 18.0$, fail to reveal the latter population (Aungwerojwit \& G\"ansicke 
2009). Similarly, the CVs in the Cal\'an-Tololo survey, which reaches down 
only to $B_J \sim 18.5$, have orbital periods that resemble those found in 
early surveys (Augusteijn et al.~2010). 

     The CRTS provides an independent source of faint CVs. Its requirement 
for a brightness range $> 2$ mag, together with its lower magnitude limit of 
$V \sim 20.5$, and the normal dwarf nova ranges of 2 -- 5 mag and the rare 
SU UMa which range up to 8 mag, means that effectively quiescent magnitudes 
are sampled from $V \sim 20.5$ to 26, but they are probably biased towards 
the upper end of the luminosities as that is where the outburst frequency 
should be largest.

In Fig.~\ref{cssdistro} we show the on-sky distribution 
of all the $\sim 640$ CVs identified in the CRTS in an equatorial
aitoff projection; the CRTS avoids a zone of $\pm 10^{\circ}$ around the 
Galactic Plane and generally finds CVs between declinations
of $+50^{\circ}$ and $-70^{\circ}$.
All CVs are shown
as small (black) dots in this figure, with additional symbols for the 100 CVs that have
measured superhump or orbital periods in the literature (red filled squares), 
those that are observable with the 40-in and 74-in reflectors from the Sutherland
station of the SAAO (blue filled circles; declinations south of $+10^{\circ}$) and the 15 CVs with new 
periods determined in the course of the UCT CCD CV survey (black large filled circles).

\begin{figure}
\centerline{\hbox{\psfig{figure=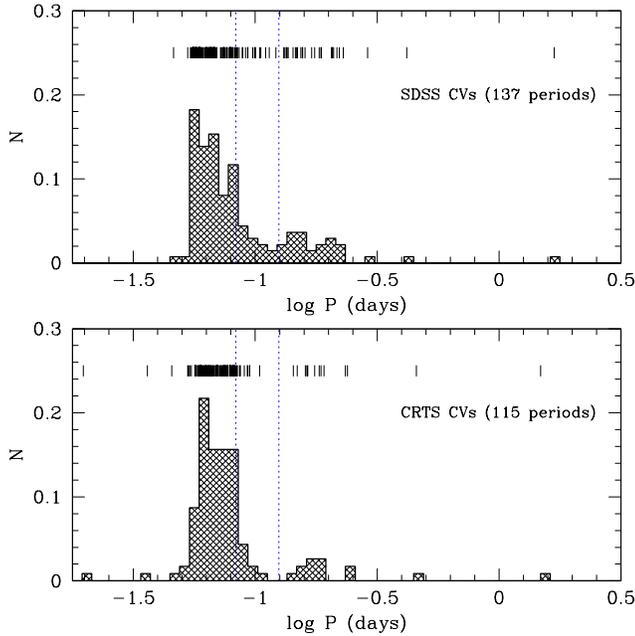,width=8.4cm}}}
  \caption{The period distribution of the 115 CRTS CVs (lower panel), compared
to distribution of 137 SDSS CVs displayed in the upper panel (G\"ansicke et al.~(2009).
The distributions are normalised by the total number of CVs in each sample. The
period gap is marked by the vertical dotted lines.}
 \label{cssfig2}
\end{figure}

We have generated a $P_{orb}$ histogram using the CRTS source catalogue 
and all previous $P_{orb}$ determinations (or $P_s$ values, converted to 
$P_{orb}$ with the G\"ansicke et al.~(2009) formula) supplemented with the 
new measurements from Tab.~\ref{survey7tab3}. Orbital periods are available
for 115 of the $\sim 650$ known CRTS CVs, with 100 systems recorded in the literature
(either in the Ritter \& Kolb (2003) catalogue, or in 
vsnet-alerts\footnote{\tt http://ooruri.kusastro.kyoto-u.ac.jp/pipermail/\newline vsnet-alert/}, where the 
latter predominantly report superhump periods) and an additional 15 orbital periods 
derived from the UCT CCD CV survey. 
Although the sample contains a small number of SDSS CVs that were already known 
before the CRTS survey began, it is almost independent and represents CVs found 
purely by the CRTS search technique. The result is shown in the lower panel of
Fig.~\ref{cssfig2}, where the SDSS distribution as found by G\"ansicke et al.~(2009) 
from 137 systems is also displayed (top panel). The CRTS histogram is similarly peaked
but narrower than the SDSS histogram below the orbital period gap; overall both distributions
reveal similar percentages of CVs in the 1.29 -- 1.55 h period bins, namely 30\% (CRTS)
versus 32\% (SDSS).

   According to current understanding of the evolution of CVs those with $P_{orb}$ 
between the minimum (‘bounce’) period and $\sim 2$ h should be a mixture of 
pre- and post- bounce systems, the latter having degenerate secondaries and 
very low $\dot{M}$. Evidence for this is slowly increasing: Littlefair et al.~(2008) 
have found three CVs where the secondaries are brown dwarfs, Patterson (2011) 
has given an extensive discussion and a list of possible candidates. This 
prediction is based on gravitational radiation (GR) being the driving mechanism 
for orbital evolution at short periods. Although that may be the only way that 
orbital angular momentum is lost on long time scales, it is clear that there is 
at least one other parameter that determines the present $\dot{M}$ in CVs: the 
spread in absolute magnitudes of CVs below the orbital period gap (see, e.g., 
figure 12 of Warner 1987) shows that $\dot{M}$ ranges over more than an 
order of magnitude, and the mean rate is several times that given by GR 
(G\"ansicke et al.~2009). With the rapidly increasing number of known short 
period CVs it may become possible to investigate this more fully. We point 
out two relevant observational points:

   The ER UMa stars lie at the high $\dot{M}$ end of the range, they have 
superoutburst intervals, $T_s$, in the range 20 -- 50 d and normal 
outburst intervals, $T_n$, $\sim 5$ d; only five members are currently known. 
They have the short orbital periods typical of SU UMa stars but appear isolated 
from the normal SU UMa stars by their short outburst intervals (Kato et al.~2003). 
They almost certainly must have exceptionally large $\dot{M}$ (though additional 
parameters seem to be required - see discussion in Kato et al.); an analogy to 
the Z Cam stars at longer $P_{orb}$ has been made (Warner 1995b). 
The few known ER UMa stars are probably only a small fraction of the true 
population - they are so variable that low cadence observations can fail 
to characterise them; indeed, the available CRTS light curves do not show their 
unusual outburst properties. The situation is similar to that of the earliest 
of them to be discovered, V1159 Ori, which was known as an irregular blue 
variable since 1906 but could not be appreciated fully until the 1990s. We 
expect that higher cadence and accepted amplitude range lowered to $\sim 1$ mag 
will provide many more examples, and perhaps fill the $T_s$ gap between the 
ER UMa stars and the standard SU UMa stars.

    More generally, we can use the CRTS light curves to segregate the 
dwarf novae into outburst classes. The light curves cover more than 5 years 
in a consistent way for all objects - they use all clear weather for 21 nights 
of each lunation and reach a cut-off magnitude $\sim 20.5$. Although outbursts 
will be missed, the same fraction will be lost independent of frequency. We have simply 
assigned class 1 to light curves showing only one outburst, class 2 to two outbursts, 
and class 3 to $>2$ outbursts. In the simplest interpretation these would correspond 
respectively to very low, low, and medium $\dot{M}$ values. 

\begin{figure}
\centerline{\hbox{\psfig{figure=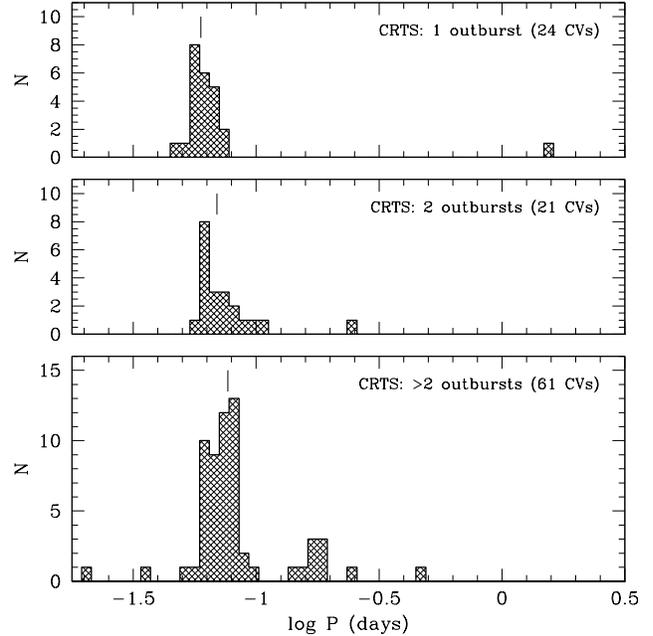,width=8.4cm}}}
  \caption{The period distribution of 106 CRTS dwarf novae: the top panel shows 
those with a single recorded outburst, the middle panel displays the dwarf novae
with two outburst measured to date and the lower panel shows those with more than
2 outbursts in five years of CRTS observations. The median value of each period distribution
is marked and corresponds to 1.43 h (top panel), 1.66 h (middle panel) and 1.82 h (lower
panel), respectively. The ultra-compact helium-rich binaries with 
$P_{orb} \le 65$ min and CVs with periods longer than 1 day were excluded from the median
statistics.}
 \label{cssfig3}
\end{figure}

   The $P_{orb}$ histograms of the three classes are shown in Fig.~\ref{cssfig3}, 
excluding nine confirmed polars in the sample of 115 CRTS CVs with observed periods. 
There is a marked correlation between $P_{orb}$ distribution and outburst class - in 
the sense that lower outburst frequencies are shifted towards shorter $P_{orb}$. But 
this trend should not arise from the simplest expected $\dot{M}$ behaviour - if GR 
is operating alone then nearly constant $\dot{M}$ is predicted (see equation 9.20 
and figure 9.2 of Warner (1995a)) and the time taken to fill dwarf nova 
accretion discs between outbursts would be independent of $P_{orb}$. Other 
factors need to be taken into account, however - the space density of CVs 
per unit interval of $P_{orb}$ is inversely proportional to the long term mean 
of $\dot{M}$ (which would cancel with outburst frequency being directly 
proportional to $\dot{M}$ to leave no $P_{orb}$ dependence on average), but outburst 
frequency is determined for each system by its current $\dot{M}$ - and as 
mentioned above this can be very different from the GR value and may depend 
on the past history of each system (e.g. when last it was a nova, in the hibernation 
scheme of things - see section 9.4.3 of Warner (1995a)). 

   A final conclusion is that the predictions of CV population studies are now 
beginning to be confirmed by observation - it required surveys reaching fainter 
than magnitude $\sim 18.5$ to detect the large population of highly evolved systems. 
Therefore the predicted end points of CV evolution should also be taken 
seriously - large numbers of cool Earth-sized white dwarfs with Jupiter-sized 
secondaries in $\sim 2$ h orbits, about 20\% of which should show total eclipses.

\section*{Acknowledgments}

PAW and BW acknowledge support from the University of Cape Town (UCT) and from the 
National Research Foundation of South Africa. PAW furthermore acknowledges financial
support from the World University Network. DdB thanks the
South African Square Kilometer Array project and UCT for financial support. SM received
support from the National Astrophysics and Space Science Programme.
We kindly thank Denise Dale for observations of SDSS0919+08.
This paper uses observations made at the South African Astronomical Observatory (SAAO).

\end{document}